\documentclass[1p]{elsarticle}
\usepackage{graphicx}
\usepackage[cmex10]{amsmath}
\usepackage{algorithmic ,algorithm}
\usepackage{underscore}
\usepackage{url}
\usepackage{lipsum}
\usepackage{array,makecell}
\usepackage{mdwmath}
\usepackage{mdwtab}
\usepackage{multirow}
\usepackage{color}
\usepackage{times}
\usepackage{amssymb}

\journal{JNCA}

\begin{document}

\begin{frontmatter}

\title{Applying Machine Learning Techniques for Caching in Next-Generation Edge Networks: A Comprehensive Survey}


\author[a1,a2]{Junaid Shuja}
\author[a1]{Kashif Bilal}
\author[a2]{Waleed Alasmary}
\author[a3]{Hassan Sinky}
\author[a4]{Eisa Alanazi}


\address[a1]{Department of Computer Science, COMSATS University Islamabad, Abbottabad Campus, Pakistan}
\address[a2]{Department of Computer Engineering, Umm Al-Qura University, Makkah, Saudi Arabia}
\address[a3]{College of Computer and Information Systems, Umm Al-Qura University, Makkah, Saudi Arabia}
\address[a4]{Department of Computer Science, Umm Al-Qura University, Makkah, Saudi Arabia}

\begin{abstract}
Edge networking is a complex and dynamic computing paradigm that aims to push cloud resources closer to the end user improving responsiveness and reducing backhaul traffic. User mobility, preferences, and content popularity are the dominant dynamic features of edge networks. Temporal and social features of content, such as the number of views and likes are leveraged to estimate the popularity of content from a global perspective. However, such estimates should not be mapped to an edge network with particular social and geographic characteristics. In next generation edge networks, i.e., 5G and beyond 5G, machine learning techniques can be applied to predict content popularity based on user preferences, cluster users based on similar content interests, and optimize cache placement and replacement strategies provided a set of constraints and predictions about the state of the network. These applications of machine learning can help identify relevant content for an edge network. This article investigates the application of machine learning techniques for in-network caching in edge networks. We survey recent state-of-the-art literature and formulate a comprehensive taxonomy based on \textbf{(a)} machine learning technique (method, objective, and features), \textbf{(b)} caching strategy (policy, location, and replacement), and \textbf{(c)} edge network (type and delivery strategy). A comparative analysis of the state-of-the-art literature is presented with respect to the parameters identified in the taxonomy. Moreover, we debate research challenges and future directions for optimal caching decisions and the application of machine learning in edge networks. 
\end{abstract}

\begin{keyword}
Caching, Edge networks, Machine learning, popularity prediction, 5G.
\end{keyword}

\end{frontmatter}


\section{Introduction}
Modern mobile applications demand low latency, mobility support, energy efficiency, and high bandwidth from back-end data stores that are usually hosted in cloud data centers and content delivery networks (CDN). The distance between mobile users and geographically distributed cloud data centers/CDNs substantially impacts latency. Future 5G and 6G networks aim to provide high-speed access to end devices while increasing traffic load of transit networks and workload of cloud data centers~\cite{Monserrat2015,Ren2019a}. CDNs and micro-data centers have been utilized to distribute the content with respect to geographic regions, hence, partially alleviating the issue of latency~\cite{Baccour2019}. It is estimated that mobile data traffic will increase 500-folds in the next decade with 80\% of the traffic contributed by video content~\cite{Index2017}. Consequently, several networking technologies have been proposed and deployed to bring computing and cache capabilities further closer to the end users. These technologies include Mobile Edge Computing (MEC), fog computing, cloudlets, and information-centric networks, etc~\cite{Ren2019a,Li2018,Qazi2019}. In general terms, the technologies providing cache and computing service in user proximity form an edge network~\cite{Liu2018,Bilal2018}. Edge network cache refers to an in-network storage that contains the content request by or relevant to edge users. The edge networks have solved the problems of high content latency in the last decade while reducing the load on backhaul networks with the help of in-network caching. The key research problems regarding caching at the edge networks are: \textbf{(a)} what (popular, multimedia), \textbf{(b)} when (time-series prediction), \textbf{(c)} where (base stations, mobile users), and \textbf{(d)} how (proactive, reactive) to cache~\cite{Sun2019}. 

5G networks and Machine learning (ML) techniques play a vital role in optimizing edge cache performance. Mobile communication technologies have experienced exponential growth over the last few years in the number of users, applications, and data traffic with services ranging from multimedia streaming to online gaming~\cite{Bilal2018,Shuja2016}. It has been identified that existing network architectures will struggle to handle the massive delivery of data efficiently with minimum latency requirements~\cite{Shafi2017}. To this end, next-generation networks (NGN), i.e., 5G evolved with the help of emerging communication technologies, such as millimeter-wave (mmWave) communication and massive Multiple Input Multiple Output (MIMO)~\cite{Bogale2016,Sim2017}. Users in 5G networks often have similar interests, and content can be stored locally (in-network cache) to reduce access latency and load on backhaul, core, and transit networks~\cite{Ren2019}. Without caching, a large amount of backhaul traffic is consumed by fetching redundant copies of popular content~\cite{Yao2019}. 5G network architectures envision caching content at various layers of access, core, and aggregate networks with the help of edge network technologies~\cite{Sutton2018}. Edge caching is one of the key elements that help 5G networks meet stringent latency requirements~\cite{Taleb2017,Elazhary2019}.   

ML techniques have been prominently used to efficiently solve various computer science problems ranging from bio-informatics to signal and image processing~\cite{Camacho2018,AliHumayun2019}. ML is based on the premise that an intelligent machine should be able to learn and adapt from its environment based on its experiences without the need for explicit programming. ML techniques are beneficial for tasks requiring classifications, clustering, decision making, and prediction~\cite{Wang2020a}. ML techniques have been adopted in wireless networks to solve problems related to, but not limited to, resource management, routing, traffic prediction, and congestion control~\cite{Sun2019,Boutaba2018,Wang2020}. Main reasons behind the emergence of ML techniques in wireless edge networks include \textbf{(a)} increase in computing resources of edge devices facilitating in-network intelligence capabilities, \textbf{(b)} generation of large amounts of data from billions of Internet of Things (IoT) devices that can be statistically exploited for the training of data-driven ML techniques, \textbf{(c)} introduction of network function virtualization (NFV) and Software Defined Networks (SDN) for programmability and manageability of network devices, and \textbf{(d)} advent of collaborative and federated learning techniques for workload distribution of compute-intensive ML algorithms~\cite{Chen2019,Jiang2016,Plastiras2018}. 

One of the applications of machine learning in wireless networks is to estimate future user requests based on mobility, popularity, and preference data sets that have time-series dynamic characteristics. Several factors challenge the process of learning user characteristics in edge networks. Firstly, user privacy policies prohibit the sharing of user data among applications and service providers. Therefore, it is difficult to formulate a generic application-independent caching policy. Secondly, most of the user traffic is https based limiting insights to user requests over the Internet for any application~\cite{Zhang2018a}. Thirdly, edge networks evolve with dynamic user associations, preferences, and content popularity. Therefore, the learning processes must address the requirements of time-variant data~\cite{Ozcan2019}. Fourthly, the learning process should predict trends specific to an edge network while meeting the storage constraints at the edge. ML techniques can be leveraged to solve the challenges of what, where, and when to cache in edge networks. The application of ML techniques is more suitable to these challenges as: \textbf{(a)} ML techniques learn with few or no data~\cite{Sadeghi2017}, \textbf{(b)} ML techniques, such as federated learning can be used for privacy-preserving mechanisms of user data~\cite{Liu2020}, \textbf{(c)} ML techniques can identify dynamic online communities and popular content with time-variance facilitated by data-driven analytics~\cite{Shan2019}, and \textbf{(d)} ML techniques can optimize a cache placement and delivery problem for a set of network states and storage constraints~\cite{Li2018b}. Given the constraints on the availability of user data in edge networks, ML models that learn without prior knowledge are often applied to optimize caching decisions~\cite{Lien2018,Lim2020}. 

\begin{table*}[htbp]
\centering
  \caption{Comparison with existing surveys}
	\label{tab:ex}
		\resizebox{.99\textwidth}{!}{

  \begin{tabular}{|m{0.8cm}|m{4.5cm}|m{4.5cm}|m{4.5cm}|} \hline
	  \textbf{Ref.} & \textbf{Contribution} & \textbf{ML Techniques Covered} & \textbf{Concentration}  \\ \hline
	~\cite{Wang2020} & Convergence of DL and edge computing & Deep Learning & Overall edge computing paradigm  \\ \hline
	~\cite{Lim2020} & Application of FL in MEC & Federated Learning & Overall edge network management  \\ \hline
  ~\cite{Sun2019} & Application of ML in wireless networks & Machine Learning & Overall wireless network management  \\  \hline  
	~\cite{Chen2019} & ANN based wireless network management & Artificial Neural Networks & Overall wireless network resource management \\  \hline
	~\cite{Yao2019} & Survey of mobile edge caching & No & Edge caching (criteria, locations, schemes, and process)  \\ \hline
	~\cite{Li2018} & Content placement and delivery strategies in cellular networks  & A brief discussion on ML techniques & Edge caching (networks, performance, and strategies) \\ \hline
	This study & Application of ML for in-network edge caching  & Comprehensive ML techniques (supervised, un-supervised, RL, NN, and TL) & Edge caching (cache locations, policy, replacement, and networks)  \\ \hline
	 \end{tabular}}
\end{table*}

\subsection{Motivation}
Modern mobile devices execute sophisticated applications, such as voice recognition, augmented reality, and high definition multimedia~\cite{Lien2018,Shuja2018,Hemanth2020}. Such applications require Ultra-Reliable Low Latency Communications (URLLC) and high user QoE. Next-generation 5G technologies envision to support the requirements of modern applications with disruptive data communication technologies (for example, mmWave communication~\cite{Bogale2016}), caching at various layers of the network~\cite{Ren2019}, and edge network architecture~\cite{Zahoor2020}. Therefore, to cope with ever-increasing bandwidth demands of emerging applications and constrained transit, backhaul, fronthaul, and radio access networks (RAN), intelligent user-centric in-network edge caching is indispensable~\cite{Taleb2017,Sutton2018}. The research challenges of cached edge networks due to their time-variant characteristics can be extensively solved with the application of ML techniques~\cite{Ozcan2019,Wang2020}. ML techniques can infer what to cache while driven by data from the user's historical content request~\cite{Wang2017a}. ML techniques can learn where to cache by predicting user locations from location histories and current context~\cite{Shan2019}. Similarly, ML techniques can optimize the caching decisions while contemplating network state and constraints of D2D and mmWave communication~\cite{Yin2018a,Chen2017}. Moreover, ML techniques can learn time-series content popularity from user request patterns and content features~\cite{Song2019}. The learning process can execute at any tier of the network including mobile devices, edges, and cloud servers, and can be federated or distributed~\cite{Yu2018,Lim2020}. Therefore, the application of ML techniques in intelligent and optimal edge caching decisions is essential to meet the requirements of multimedia and URLLC content delivery~\cite{Ahmed2017,Azimi2018}. ML techniques employed towards optimal edge caching decisions are driving technologies and use cases with low latency requirements, e.g., autonomous vehicles, online gaming, virtual and augmented reality, dynamic content delivery, video streaming, IoT, and smart cities~\cite{Tapwal2020,Chekired2019,Wang2020,Khan2020,Wang2017}. A survey specifically focusing and detailing the role, opportunities, and challenges of ML applications towards caching in edge networks was missing in the literature. We aim to fill the missing gap by specifically focusing on ML and caching in edge networks.

\subsection{Existing surveys}
In this article, we extensively survey ML-based edge caching techniques. Significant progress has been made in surveying applications of ML techniques in the edge, wireless, and cellular networks. Earlier surveys either focused on edge caching without focusing on ML techniques~\cite{Yao2019,Li2018,Goian2019,Wang2017}, or focused on the application of ML techniques to overall resource management in wireless networks~\cite{Sun2019,Lim2020,Wang2020,Chen2019}. Zhu et al.~\cite{Zhu2018} survey the application of Deep Reinforcement Learning (DRL) towards mobile edge caching. However, the survey was brief and focused only on the application of a specific ML class (DRL) to edge caching. Wang et al.~\cite{Wang2020} provided a comprehensive survey on deep learning applications and inferences towards resource management of edge computing paradigm. Researchers in~\cite{Lim2020} surveyed federated learning-based edge networks while focusing on resource management and privacy issues. Sun et al.~\cite{Sun2019} provided a comprehensive review of ML applications in wireless networks including resource management, backhaul management, beamforming, and caching. A comprehensive survey of cellular cache optimization techniques (stochastic, game theory, graph-based, and ML) was provided in~\cite{Yao2019}.  

This survey can be distinguished from earlier surveys on caching at the edge based on the fact that: \textbf{(a)} it focuses on the comprehensive application of ML techniques rather than a specific ML method~\cite{Zhu2018,Wang2020,Lim2020} and \textbf{(b)} it focuses on optimal caching decisions in edge networks rather than overall resource management~\cite{Chen2019,Sun2019}. Table~\ref{tab:ex} further lists existing surveys in the same direction to highlight the differences. 


\subsection{Contributions and organization}
The main contributions of this article are as follows:
\begin{itemize}
\item We formulate a taxonomy of the research domain while identifying key characteristics of applied ML techniques, edge networks, and caching strategies.
\item We detail the fundamentals of the research domain of edge networks while discussing the role of 5G and network virtualization in edge caching.
\item We enlist state-of-the-art research on ML-based edge caching while categorizing the works on ML techniques, i.e., supervised learning, unsupervised learning, neural networks (NN), reinforcement learning (RL), and transfer learning (TL). We compare state-of-the-art research based on the parameters identified in the taxonomy to debate the pros and cons.
\item We comprehensively debate key research challenges and future directions for the adoption of ML techniques for caching at the edge networks ranging from federated caching, content transcoding, user privacy, and federated learning.
\end{itemize}

\begin{table*}	[htbp]
\centering
  \caption{List of abbreviations} 
	\label{tab:ab}}{
	\resizebox{.99\textwidth}{!}{
  \begin{tabular}{|m{1cm}|m{4cm}|m{1cm}|m{4cm}|m{1.2cm}|m{4cm}|} \hline
	\textbf{Abbr.} & \textbf{Full Form} & \textbf{Abbr.} & \textbf{Full Form} & \textbf{Abbr.} & \textbf{Full Form}   \\ \hline
    {5G} &  Fifth generation & {AP} & Access Point & {API} & Application Programmer Interface \\\hline
		{BS} & Base Station & {BBU} & Baseband Unit & 		{CDN} & Content Delivery Network\\\hline
		{CHR} & Cache Hit Ratio & {CNN} & Convolutional Neural Network & 	{CoMP} & Coordinated MultiPoint \\\hline
		{C-RAN} & Cloud Radio Area Network & {D2D} & Device to Device &	D(RL) & (Deep) Reinforcement Learning \\\hline 
		FL & Federated Learning & {F-RAN} & Fog Radio Area Network & HTTP & HyperText Transfer Protocol \\\hline
		HetNet & Heterogeneous Networks &	{IoT} & Internet of Things &	{MBS} & Macro Base Station \\\hline
		{MEC}  & Mobile Edge Computing & {MDP} & Markov Decision Process & mmWave & millimeter-wave \\\hline
	  {MIMO} & Multiple Input Multiple Output & {MNO} & Mobile Network Operator &	{ML} & Machine Learning  \\\hline
		{NGN} & Next Generation Networks  & {NN} & Neural Network &	{NFV} & Network Function Virtualization \\\hline
		{QoE} & Quality of Experience & {QoS} & Quality of Service & RAN & Radio Access Network \\\hline
		{RNN} & Recurrent Neural Network & {RRH} & Radio Remote Heads & {SBS} & Small Base Station \\\hline
		SDN &  Software Defined Networks & {SVM} & Support Vector Machine & {TL} & Transfer Learning \\\hline 
		{UE} & User Equipment  & {URLLC} & Ultra-Reliable Low Latency Communications  & {V2I} & Vehicle-to-infrastructure \\\hline

  \end{tabular}}
\end{table*}
The rest of the paper is structured as follows. Section~\ref{sec:funda} presents the fundamentals of edge networks, 5G technologies, and network virtualization. The  discussion includes the role of caching towards 5G networks. Section~\ref{sec:tax} lists the taxonomy of ML techniques for caching in edge networks. In Section~\ref{sec:state} and~\ref{sec:comparison}, we present state-of-the-art caching techniques based on ML and their comparison. Key challenges and future research directions are comprehensively discussed in Section~\ref{sec:cfd}. Section~\ref{sec:conc}, concludes the article. 

\section{Fundamentals of Edge Networks and 5G Technologies}\label{sec:funda}
Edge caching defines a edge user-centric cache that has limited storage capacity and saves content that are only relevant to the edge users~\cite{Wang2017,Tran2017}. The limited storage capacity of edge and the dynamic characteristics of user (mobility, content access, etc) make the caching decision distinct from CDN caching. The optimal caching decisions in edge networks are based on multiple inputs such as, content popularity prediction, user mobility prediction, user access prediction, and wireless channel condition prediction. As a result, ML techniques are employed to predict the input features to the optimal caching problems~\cite{Sun2019,Chen2019}. We provide an overview of edge networks, 5G, and network virtualization in this section. We also discuss the role of edge caching in 5G networks. Moreover, the role of network virtualization technologies (SDN and NFV) in edge caching is explored. The list of abbreviations adopted in this article are listed in Table~\ref{tab:ab} for better article readability.  

\begin{figure*}
\centering
\includegraphics[height=4.5cm]{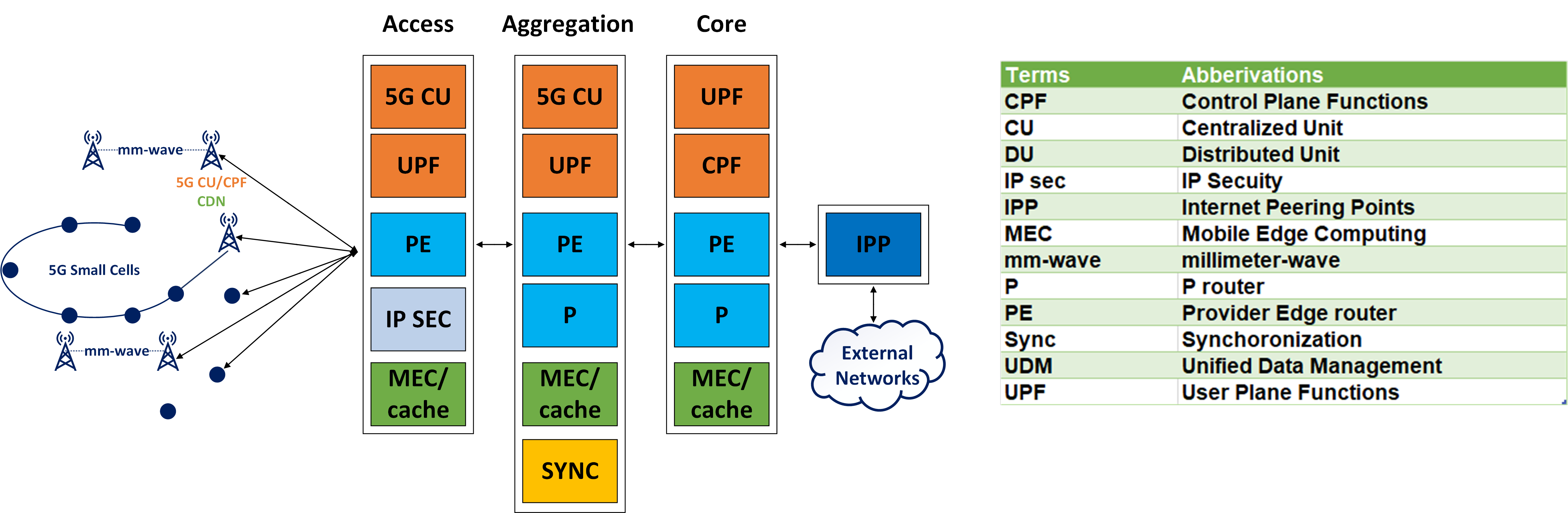}
\caption {5G conceptual architecture~\cite{Sutton2018}}
\label{fig:5g}
\end{figure*}
\subsection{Edge Networks}
The principle of edge computing is to relocate compute, storage (cache), and communication services from centralized cloud servers to distributed nodes (users, BSs) located at the edge of the network~\cite{Wang2017,Kibria2018}. The inter-connected edge resources form an edge network. The principle of edge computing is supported by a set of networking technologies that vary in architecture and protocol. Edge computing technologies include cloudlets~\cite{Jararweh2016}, fog computing~\cite{Bilal2018}, and MEC~\cite{Ahmed2017,Othman2017}. While locating resources at the edge of the network in user proximity, edge computing promises reduced service latency, backhaul bandwidth reduction, UE energy efficiency, and context-awareness that can be utilized for intelligent service placement.     

MEC was envisioned by the European Telecommunications Standards Institute (ETSI) proposing virtualization of resources to enable execution of services at the network edge~\cite{Hu2015}. Hence, network virtualization is a key enabler for MEC. The architecture of fog computing was proposed by Cisco to extend cloud services to the wireless edge network specifically serving IoTs~\cite{Klas2015}. Fog computing deploys an intermediate platform between end devices and cloud infrastructure to distribute services near users and facilitate time-sensitive applications~\cite{Bilal2018}. Cloudlets represent the middle tier of the three-tier architecture consisting of mobile devices, microdata centers (cloudlets), and the cloud~\cite{Satyanarayanan2009}. The key enabler for cloudlets are also virtualization technologies as virtual images representing cloud services are moved towards the edge networks~\cite{Elazhary2019,Shuja2016a}. Edge networks can be defined as a network architecture that deploys flexible computing and storage resources at the network edge, including the RAN, edge routers, gateways, and mobile devices, etc., with the help of SDN and NFV technologies~\cite{Wang2017,Ahmed2017}. Edge caching considered in this article can encompass any of the above-mentioned edge computing technologies.  

\subsection{Role of caching in 5G technologies}
The idea of caching was first applied to Internet services in the form of CDNs. The content from cloud data centers is stored in CDNs which are distributed over a region that may encompass multiple states and countries. Edge network caching brings the content storage further near to the end user with support of enterprise networks and network service providers. Edge caching has been considered recently to improve content delivery in wireless networks~\cite{Bilal2018,Ahmed2017}. The role of caching in NGNs can be emphasized by putting forward a question: Can edge caching tackle the challenges of URLLC, QoE, and cost in 5G networks? The answer is an emphatic yes. The advancements in 5G networks and high-resolution multimedia content means that the network bottleneck will shift to the core and transit links~\cite{Paschos2018}. Therefore, caching in user proximity may be the only viable solution for growing user requirements from multimedia and virtual/augmented reality applications with time-sensitive stipulations~\cite{Yao2019,Schwab2020}.

With the rapid advancements in mobile communication technologies, an ever-increasing number of mobile users are experiencing a wide variety of multimedia services, such as video sharing, video conferencing, real-time video streaming, and online gaming using smartphones and tablets. According to Cisco, the global mobile data traffic is 41 Exabytes per month, and it will increase to 77 Exabytes, containing 79\% of video data, by 2022~\cite{Index2017}. This exponential growth had put a huge burden on the capacities of RAN, transit link, and core network due to the centralized nature of mobile network architectures~\cite{Wang2015,Li2015,Li2015a}. Research and academia identified some major requirements for NGNs, i.e., deliver the content with minimum latency, provide higher bandwidth, and support 1000 times more number of users~\cite{Monserrat2015,Shafi2017,Andrews2014}. Specifically, the current mobile network architectures are unable to efficiently handle the huge delivery of contents that are repeatedly requested by multiple users, for example, the request for the same Ultra High Definition (UHD) videos put a huge burden over the transit link and core network. The massive and variable traffic demands make it difficult to achieve good service performance and deliver high QoE for user-centric applications~\cite{Paschos2018,Chen2014}.

To address the above-mentioned challenges, novel mobile architectures and advanced data communication technologies, such as mmWave communication~\cite{Taori2015}, MIMO)~\cite{Xiang2014}, ultra-wideband communication~\cite{Zhang2019}, and Heterogeneous Networks (HetNets)~\cite{Bogale2016} are evolving the advancement of next-generation, i.e., fifth-generation (5G) mobile networks. 5G cellular networks system designs in collaboration with new application-aware approaches improve the network resources utilization and reduce traffic demands. A generic conceptual architecture of 5G developed by Sutton et al.\cite{Sutton2018} is illustrated in Figure~\ref{fig:5g}. The figure depicts the proposed architecture for the 5G networks. Edge computing and caching is an integral part of 5G network and holds a key role in attaining the 5G objectives. The edge commuting support and caching can be placed at any layer (core, aggregation, and access) and even be situated at base stations in the proposed 5G architecture. The flexibility for edge placement is offered to ensure the appropriate and optimal use of constrained edge resources as desired. 

One of the major approaches to reduce the burden from the backhaul and core network is to cache the popular content in the near vicinity of the users to reduce the redundant downloads of similar content~\cite{Wang2014,Woo2013,Rodrigues2020}. For this purpose, the Content Delivery Networks (CDNs) caches are proposed as an integral part of 5G network architecture. According to Andy Sutton, a principal network architect at BT, the conceptual 5G network architecture, as shown in Fig~\ref{fig:5g}, provides the caching facility at different layers of the hierarchical model~\cite{Sutton2018,Khan2019}. According to architecture, the CDNs can be deployed at the Core, Aggregation, Access, and even at Base Station (BS) layer. By leveraging the content placement strategies, popular content can be cached proactively during the off peak-hours to reduce the burden on the transit link during the peak hours~\cite{Mueller2016,Qiao2016}.

The requirements of URLLC for NGNs are driving the cache locations further near to the users. The emerging paradigm in edge caching is caching at UE with D2D communication for content diffusion. The URLLC UE caching architecture arises from the elements of \textbf{(a)} high-speed D2D communication protocols, \textbf{(b)} high-storage capability in modern smartphones, and \textbf{(c)} emerging ultra-dense networks~\cite{Shen2017,Li2019a,Association2016}. As a result, caching strategies are migrating from centralized CDNs to highly distributed UEs. The UEs are highly mobile, have individual content preferences, communicate in social-aware patterns, and are part of multiple online social communities. The ML techniques come to the forefront to solve complex problems of popularity and mobility prediction, user and content clustering, and social community detection in edge networks~\cite{Nitti2019,Shan2019}. 

\subsection{Role of SDN and NFV in Edge Caching}
The SDN/NFV technologies are gaining momentum and are enablers of intelligent edge and cloud networks~\cite{Paschos2018}. SDNs are based on the principle of decoupling data plane from the control plane of the network~\cite{Huo2016}. As a result of the decoupling, SDNs provide three key characteristics of flexibility, programmability, and centralized control for ML-based edge networks. The network functions can be abstracted and programmed for evolving application requirements and logically implemented through a centralized control plane throughout an edge network~\cite{He2017a}. The benefits of implementing SDNs for ML-based edge caching are three-fold. First, the programmable control plan can be utilized to implement a distributed in-network intelligence framework among network entities~\cite{Lei2018,Din2019}. Secondly, ML techniques are data-driven. The SDN controller has a centralized view with the ability to collect network data and enable the application of ML techniques~\cite{Xie2019}. Thirdly, SDNs allow flexible routing of user requests to distributed and hierarchical cache locations within the network~\cite{Mehrabi2019}. 

NFV allows virtualization of the common physical channel into multiple virtual channels/functions that can be shared among users with varying QoS requirements. Moreover, NFV decouples network functions from dedicated network devices so that they can be implemented over general-purpose computing systems~\cite{Yousaf2017,OrdonezLucena2017}. As a result, in-network intelligence and network functions can execute on the same general-purpose computing platform. The network virtualization technologies are creating further opportunities for edge caching while empowering MNOs to manage the compute, storage, and network resources. The virtualization technologies are helping in the migration of caching services from CDNs to the wireless edge. The caching service can be implemented as a Virtual Network Function (VNF) at the network edge with the flexibility of cache initiation and scaling on the fly. Allocating in-network edge resources falls under the category of network slicing~\cite{OrdonezLucena2017,Mehrabi2019}. More importantly, network virtualization technologies enable the implementation of learning capabilities at the edge network, hence, leading to optimal caching decisions. Moreover, available free resources at any network layer may also be allocated elastically as required according to the QoS requirements. Therefore, it can be established that network virtualization technologies are empowering 5G networks with edge caching and intelligence capabilities~\cite{Yousaf2017,Taleb2016}. ML techniques have also been extensively applied to SDN technologies for resource management, security, and routing optimization~\cite{Xie2019}. 
\begin{figure*}
\centering
\includegraphics[width=13cm]{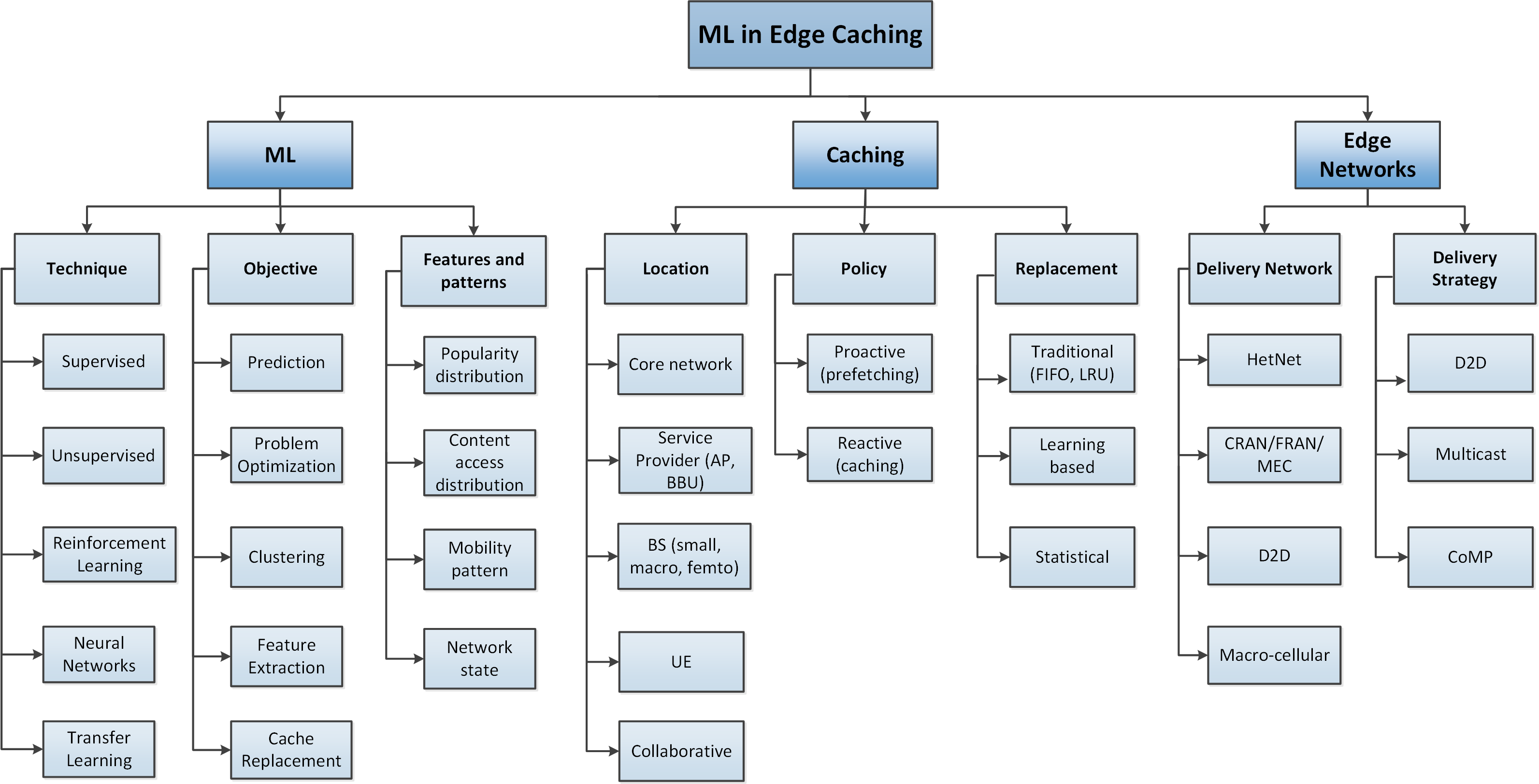}
\caption {Taxonomy of Machine Learning Techniques for Caching in Edge Networks}
\label{Fig:tax}
\end{figure*}

\section{Taxonomy of ML based edge caching}\label{sec:tax} 
This section presents the taxonomy of ML applications in edge caching. The aim of this section is to provide the reader with background knowledge of the research domain in detail with the help of categorized concepts. Figure~\ref{Fig:tax} illustrates the taxonomy of the article with major classification as: \textbf{(a)} ML (technique, objective, and features) \textbf{(b)} Caching (location, policy, and replacement), and \textbf{(c)} Edge Networks (delivery network and strategy). The necessary elements of the taxonomy are explained in detail as follows. 

\subsection{ML Technique/Method} The main application of ML in caching is to answer what and when to cache so that the caching objectives (cache hit ratios, latency, throughput, etc) are optimized. The ML techniques used to find this answer can be generally characterized into \textbf{(a)} supervised learning, \textbf{(b)} unsupervised learning, \textbf{(c)} RL, \textbf{(d)} NN, and \textbf{(e)} TL  techniques~\cite{Sun2019,Goian2019}. In some scenarios, multiple ML techniques can be applied to a problem of what to cache at the edge/wireless networks with the objective to find solutions to sub-problems, such as, popularity prediction, cache decision optimization, and user clustering. Supervised learning aims to learn a general rule for mapping input to output based on the labeled data set. Example input and corresponding outputs train the learning agent to learn the mapping of the rest of the data set~\cite{Chang2018,Miyazawa2016}. Unsupervised learning aims to learn the mapping function based on un-labeled data~\cite{Cao2019,Sun2019}. A learning agent continuously interacts with its environment to generate a mapping function based on immediate response/reward in RL~\cite{He2017a}. NNs are inspired by the structure and function of biological NNs that can learn from complex data. NN generally consist of input, hidden, and output layers~\cite{Teerapittayanon2017,Lim2020}. The goal of TL is to utilize knowledge of a specific (source) domain/task in the learning process of a target domain/task. TL avoids the cost of learning a task from scratch~\cite{Chen2019a,Hou2018}. Semi-supervised learning aided by both supervised and unsupervised learning models is also a popular ML technique but rarely utilized for caching in wireless networks~\cite{Bommaraveni2019}.  
5
\subsection{ML Objective} In most of the research works listed in the next section, ML techniques are applied to multiple sub-problems. For example, the objective of ML can be: \textbf{(a)} prediction (popularity, mobility, user preference), \textbf{(b)} problem optimization, \textbf{(c)} clustering (user, BS, content), \textbf{(d)} cache replacement, and \textbf{(e)} feature extraction~\cite{Sun2019,Wang2017}. Prediction is often the most difficult task in this scenario (edge network) where assumptions are made regarding user preferences and content history. We consider prediction as to the primary objective of ML application while other objectives as secondary. The non-standard and non-public data sets also hinder research evaluations on common grounds. For example, the data for content popularity at YouTube CDNs can be obtained which pertains to a wider geographic region than edge networks. Some studies assume this wider regional data as edge network data and predict future content popularity. Similarly, user preferences are enclosed in application layer data across the network where service providers are compliant to privacy laws. Social media publishers (Twitter, Facebook, etc) provide API based access to user data with increasing restrictions due to recent data breaches~\cite{Cadwalladr2018,Wang2019b}. ML techniques have been applied to find an optimal resource allocation solutions in cached networks, i.e., where, what, how, and when to cache. ML techniques have also been employed to reduce the time complexity of the mathematical formulation of cache optimization problems. Generally, given the input and sample optimal solutions, the algorithms learn optimal solutions for varying inputs~\cite{Sun2019}. Unsupervised clustering techniques are applied to data during pre-processing stages to group data and find similarities. Clustering partitions a data-set into communities with samples that are more similar to each other based on a similarity/distance function. The clustering of users is based on their D2D connectivity and social-awareness. BSs are clustered based on their coverage area while content clustering is based on data similarity~\cite{Liu2020}. Cache replacement strategy can be learned using an ML (RL, Q-learning, etc) such that the reward/Q-values of the problem are inferred from a cache hit ratios~\cite{Tanzil2017}.  

\subsection{Features and Patterns} ML techniques often take various features and patterns of the environment as input to train the model. In edge caching, the environment variables that require modeling or have historical data and are input to the ML framework are: \textbf{(a)} content popularity distribution, \textbf{(b)} content access/request distribution, \textbf{(c)} user mobility model, and \textbf{(d)} network state model. The content popularity is modeled mostly as zipf distributions~\cite{Liu2019}. Moreover, content popularity can be predicted with the help of ML techniques based on historical information~\cite{Tanzil2017}. User content requests are modeled as poisson point distributions~\cite{Hu2018} or zipf distributions~\cite{Bommaraveni2019}. User content requests can be estimated with the help of ML techniques based on content features~\cite{Tanzil2017}. User mobility is modeled as poisson point process (PPP)~\cite{Jiang2019a}, random walk model~\cite{Malik2020} or Markov process~\cite{Liu2019}. User mobility can also be predicted with the help of ML techniques given user context and historical information regarding user behavior~\cite{Li2019}. The dynamic wireless network state is modeled as finite-state Markov channel~\cite{Xiang2019} or assumed to be invariant. 

\subsection{Cache Placement} Cache placement answers the query of where to cache. The caching places in a wireless edge network can be \textbf{(a)} User equipment (mobile devices, home routers), \textbf{(b)} BSs (small, macro, femtocells), \textbf{(c)} BBU pool/CRAN, \textbf{(d)} mobile core network, and \textbf{(e)} collaborative as joint placement in multi-tier caching architecture~\cite{Said2018,Yao2019}. Mobile devices are becoming increasingly sophisticated with larger storage capacities and can act as in-network caches. Caching at UE is also known as infrastructure-less caching and leads to lower the burden on access/cellular networks, low network latency, and higher spectrum utilization while offloading wireless traffic. D2D links in unlicensed-band are required for content communication among UEs. BS of all form and sizes (Macro, small, femto, etc) provide relatively higher latency compared to local caching but connect to a larger set of users in the wireless coverage area. Similarly, service providers in the form of C-RAN, F-RAN, and MEC often provide proximate cache and compute services in user hotspots. Virtualized Baseband Unit pools (BBU) can be utilized as cache location for service providers with the help of SDN and NFV technologies. Service provider's independent infrastructure in the form of edge servers can be added and connected to MNO equipment for content caching~\cite{Wang2019}. Service provider caches provide higher caching capacity and area coverage with higher latency~\cite{Yao2019}. Moreover, C-RAN based cache placement enables service providers to gather vital content features (hits, likes, etc) from central or geographically federated data centers and employ them in the learning process to improve cache hit rates. Some researchers have utilized collaborative caching techniques where the cache placement is distributed among multiple elements in the wireless network with the help of coding schemes~\cite{Fadlallah2017}. The content-coding schemes divide each content into smaller segments for distributed placement and cache size optimization~\cite{Li2018}. Cache placement can also be federated and collaborated among MNOs to reduce redundancy~\cite{Taleb2019}. While caching content closer to the user, user mobility is a larger challenge due to its D2D communication link coverage from cache enabled UE and BS association. Caching content at C-RAN reduces the challenge of user mobility~\cite{Ye2019,Goian2019}. 

\subsection{Caching Policy} 
\begin{figure}
\centering
\includegraphics[width=8.5cm]{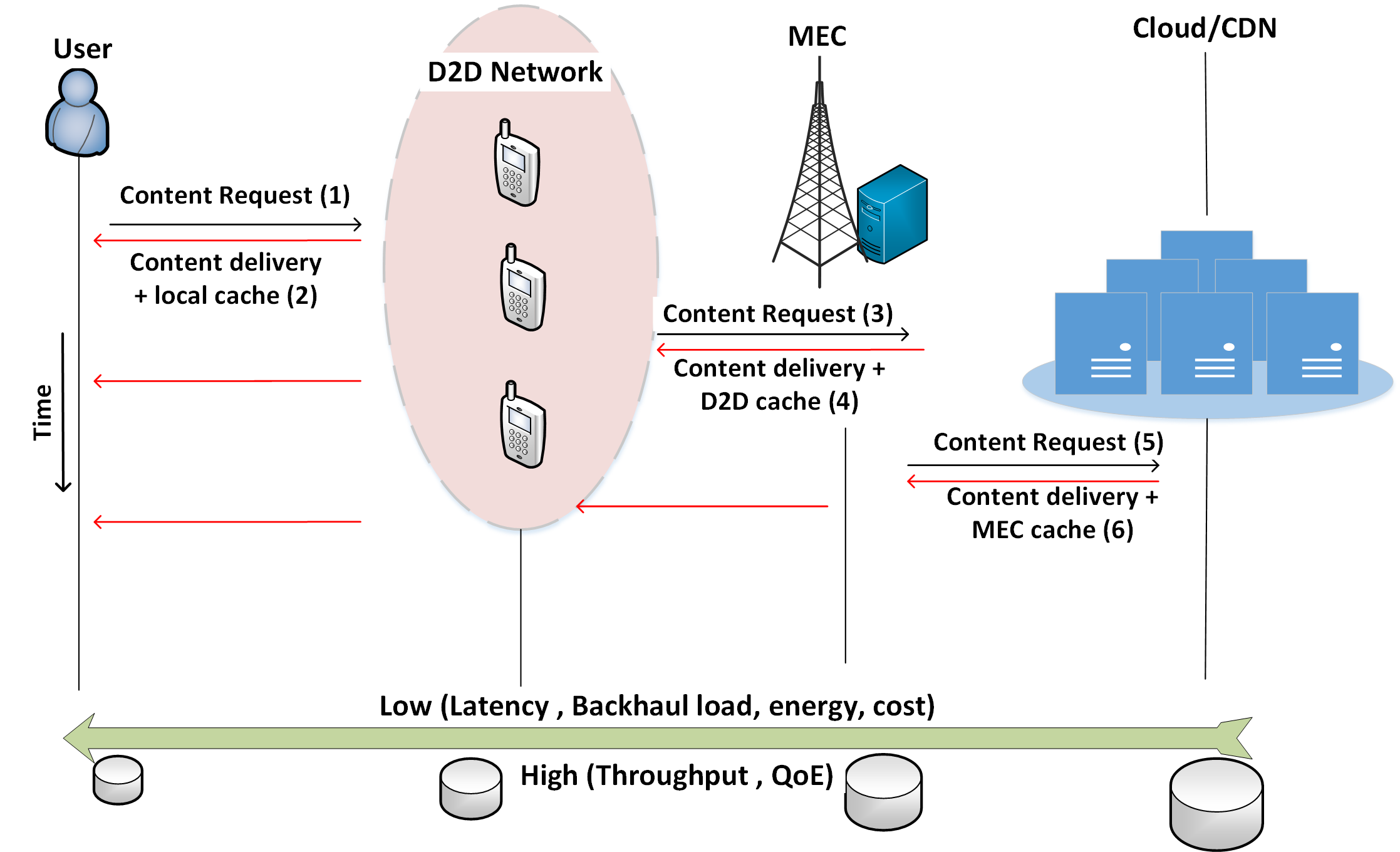}
\caption {Reactive caching in Edge Networks}
\label{fig:re}
\end{figure}

The caching policy can be either proactive or reactive and solves the problem of how to cache while partially responding to when to cache. In the reactive policy, the content is cached once it is requested by a user. When a user request a content, it is delivered from local cache if found. If the content is not found, it is progressively requested from nearby devices in a D2D network, MEC server, and CDNs. During the content delivery phase, it may be cached at multiple locations according to the caching policy. Figure~\ref{fig:re} illustrates the concept of reactive caching in Edge networks. 

\begin{figure}
\centering
\includegraphics[width=8.5cm]{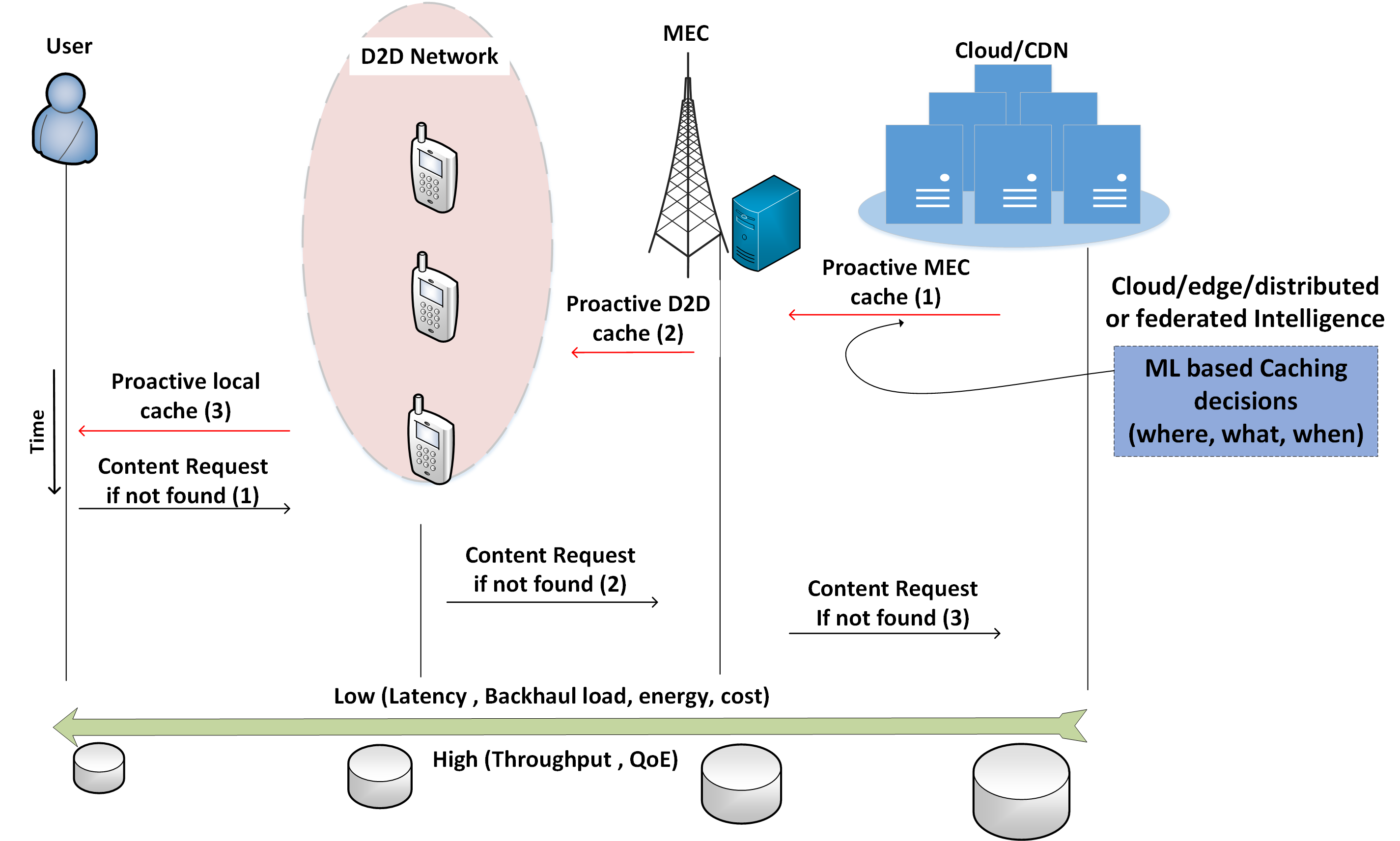}
\caption {Proactive caching in Edge Networks}
\label{fig:pro}
\end{figure}

The proactive approach predicts the popularity based on user preferences and pre-fetches content to further lower access time. A ML framework executing on a cloud or MEC server can predict content popularity based on an edge network's user access features and proactively push content to edge caches. Proactive caching policies have their benefits and challenges. Proactive caching considerably reduces the delay, increases the QoE, and relieves the burden form the transit and backhaul links, specifically in peak hours. Proactive caching can also opportunistically cache content such that network non-peak hours are utilized for UE caching leading to higher spectral efficiency~\cite{Yao2019,Chang2018}. However, proactive caching may result in poor cache hit ratio specifically in edge network where a content which may not be popular is cached on the edge. Moreover, proactively caching a content which may not be accessed requires storage space for caching, which can only be gained by removing stored content. Consequently, redundant fetches may occur for the deleted content specifically in case of small caches. Therefore, it is necessary to apply ML techniques aided by content popularity, user access, and mobility data to benefit from proactive caching and address its challenges~\cite{Liu2019,Baccour2020a}. The learning process is difficult as it increases the number of sub-problems, such as popularity prediction and mobility prediction. Figure~\ref{fig:pro} depicts the idea of proactive caching in Edge networks. 

\subsection{Cache Replacement} In general computing and CDN caching systems, Least Frequently Used (LFU), Least Recently Used (LRU), and First In First Out (FIFO) algorithms are utilized for cache replacement/eviction. However, the standard recency-based and frequency-based caching algorithms can not be applied to wireless edge networks due to its non-deterministic properties~\cite{Xu2018}. Cache replacement techniques are also necessary as the caching capacity of most network devices is very limited as compared to CDN resources. Moreover, approximately 500 hours of videos are uploaded on YouTube alone every minute~\cite{Index2017}. Therefore, considering the rate of data generation, cache population and replacement play a significant role in user QoE~\footnote{https://www.statista.com/statistics/259477/hours-of-video-uploaded-to-youtube-every-minute/}. The approach followed by some of the state-of-the-art works in wireless networks is learning-based cache replacement. Generally, a Q-learning algorithm can be applied by a UE to act as an independent or joint learner to learn the reward/Q-value of caching/replacement decisions~\cite{Jiang2019a,Wang2017a,Lee2020}. On the other hand, some research works have not considered the sub-problem of cache replacement~\cite{Chang2018} while some have considered a caching decision problem for instance \textit{t+1}~\cite{Thar2018}. 
\begin{figure*}
\centering
\includegraphics[width=13cm]{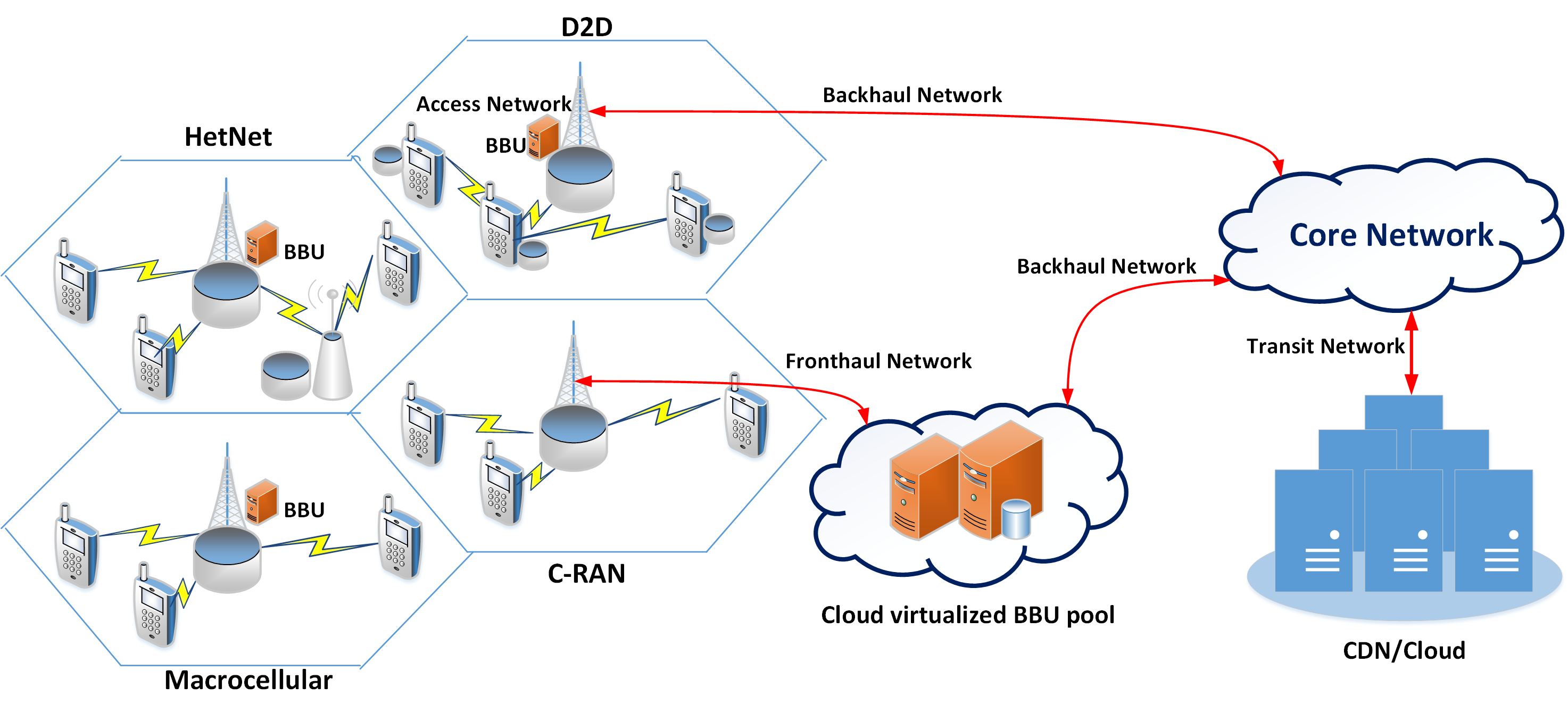}
\caption {Types of delivery networks in edge caching}
\label{fig:delivery}
\end{figure*}

\subsection{Delivery Network Type} The access layer of the edge network considering caching architectures can be categorized as \textbf{(a)} macro-cellular networks, \textbf{(b)} HetNet, \textbf{(c)} C-RAN/ F-RAN/ MEC, and \textbf{(d)} D2D networks~\cite{Li2018,Wang2017}. Figure~\ref{fig:delivery} depicts the four types of delivery networks for edge caching connected to a core network. In macro-cellular architectures, UEs are connected to a single macro BS at a time~\cite{Woo2013}. HetNets comprise of UEs with multiple access technologies, such as 5G and Wi-Fi~\cite{Mehrabi2019}. Moreover, HetNets can also be defined as networks where UEs lie in the coverage of multiple BSs (caches) of different cell sizes such as macro and pico~\cite{Zhang2015}. Content can be cached on any of the BSs. However, HetNets cache content on multiple BSs so that the UEs can obtain content from the closest BS based on its communication parameters~\cite{Kabir2020}. The caching capacity and coverage area in HetNets is generally low as compared to macro-cellular architecture. C-RAN/F-RAN cellular architectures employe a central pool of BBU resources with the help of virtualization techniques for processing and caching~\cite{Chen2018}. Cache-enabled D2D networks utilize UE caching capacity and communicate content in close range bypassing BSs and MNOs. UEs often coordinate in clusters based on short-range communication~\cite{Nitti2019}. 

\subsection{Delivery Strategy} Content delivery strategies try to reduce the load on both front-haul and back-haul networks and reduce duplicate packet transmissions. Delivery strategies employed to deliver cached content in wireless networks are \textbf{(a)} D2D, \textbf{(b)} multi-cast, and \textbf{(c)} coordinated multipoint (CoMP) transmissions. D2D communications are possible in dense networks where content cached on one user device can be sent to nearby users using the unlicensed band (Bluetooth and Wi-Fi)~\cite{Li2019}. D2D caching also utilizes relay/helper nodes when the source and destination nodes can not communicate directly with short-range communication~\cite{Yao2019}. The state of a UE battery may determine their participation in D2D communication. Subsequently, D2D caching requires incentives for UE to act as caching helpers over longer periods of time~\cite{Liu2016}. Multi-cast delivery in caching enabled networks results in packet delivery to multiple users within the network cell. Often in popular live streaming videos in online social networks and video-on-demand services, content is simultaneously watched by multiple users and multi-cast delivery results in reducing duplicate packet transmission~\cite{Bilal2019}. MNOs collect user requests for content over a time frame so that multiple requests for the same content can be furnished simultaneously. Multi-cast delivery has not been discussed in detail in edge caching scenarios. However, amalgamation of multi-cast with ML techniques can result in considerable gains in caching, RAN resource management, and user QoE. ML based prediction techniques may be applied to multi-cast same content to multiple UEs for proactive caching saving RAN resources, minimizing delay, and delivering high QoE. CoMP refers to various techniques that aim for dynamic coordination of data transmission/reception at multiple geographically separate sites~\cite{Wang2017}. In particular, a UE lying at cell edge can be connected to multiple BSs with the aim of improving overall QoE. The user can be connected to one of the BSs which does not have cache of requested content. CoMP delivery can be employed to provide the requested content from other BSs within the coverage network. A non-traditional content delivery strategy is to change the user association in a HetNet with overlapped coverage so that the user is associated with a BS which has the cached content~\cite{Liu2016}. Cache delivery strategies also optimize UE power allocation, channel allocation, and other required transmission parameters~\cite{Li2018}. 

The main objectives of caching in edge networks are \textbf{(a)} low content delay/access time \textbf{(b)} low backhaul load, \textbf{(c)} energy efficiency for both end user and MNO, \textbf{(d)} higher network throughput, \textbf{(e)} higher spectral efficiency, \textbf{(f)} reduce MNO cost, and \textbf{(g)} increase user QoE~\cite{Li2018,Wang2017}. The optimization problem can be formulated such that it reduces one or more than one of these parameters. The application of ML to caching can further optimize these parameters as a result of higher CHR~\cite{Wei2018}. Caching and backhaul data offloading in edge networks often serve the same concept and purpose~\cite{Wang2017a,Zhao2019}. We use the term of edge caching throughout this article. The user request for content can be realized from many sources. The requested content can be accessed from the local cache, nearby UEs with D2D communications, and BSs with wireless communication. The content can be cached in the MEC server, virtualized C-RAN resources, or central data stores in the form of CDN/clouds~\cite{Yao2019}. 
\section{State-of-the-art: Machine Learning based Caching}\label{sec:state}
\begin{figure*}
\centering
\includegraphics[width=13cm]{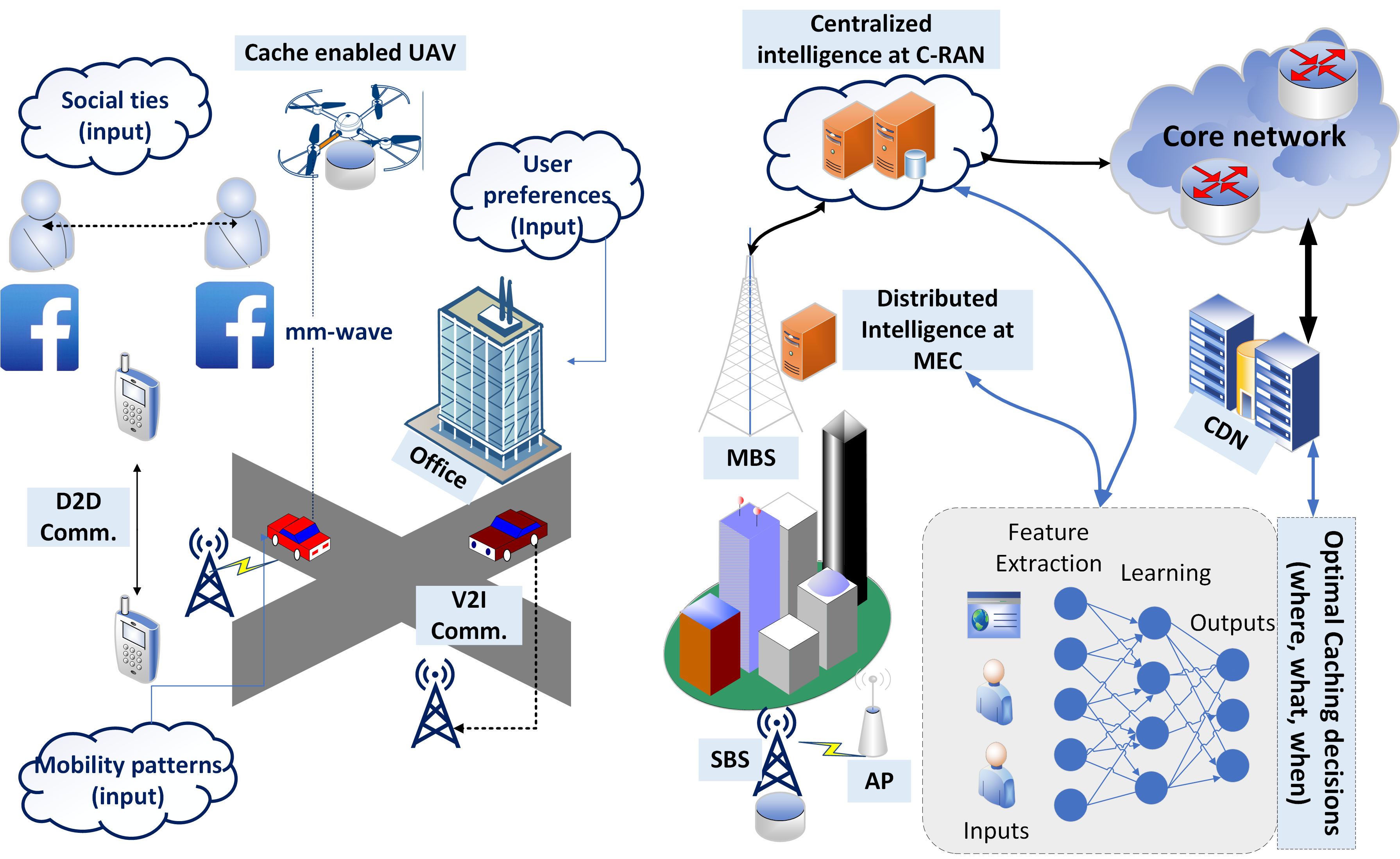} 
\caption {The Concept of Edge based Intelligence for Edge Caching}
\label{fig:int}
\end{figure*}

Figure~\ref{fig:int} depicts a high-level conceptual framework of ML-based edge caching. Social-awareness~\cite{Bui2018}, mobility patterns~\cite{Hu2018}, and user preferences~\cite{Ren2019} are input to an ML framework that can be federated among multiple nodes, hosted at distributed MEC servers, C-RAN, or centralized CDN. The inputs to the ML based caching decision framework are extracted from social networks, D2D communications, and V2I communications. The intelligent and optimal decisions are fed-back to the caches that are managed by network virtualization techniques~\cite{He2017a,Luo2019}. Innovation is brought to caching placement techniques with the help of UAV mounted caches that can be directed towards dense user populations and real-time events~\cite{Cheng2018}. The low-level design considerations of ML based edge caching will be debated and illustrated in forthcoming sections.

We divide the state-of-the-art studies regarding caching in edge networks based on ML techniques categorized into supervised, unsupervised, RL, NN, and TL based methods. Two uncategorized works are listed in a separate subsection. The discussions of remaining taxonomy branches (ML objective, cache locations, network type, etc) are incorporated within this section to avoid redundancy of the ideas considering the length of the article. Most of the related works utilize more than one ML technique where one of the ML techniques is used in prediction (popularity, mobility, user preference) while the other is utilized to find optimal solutions for caching (placement, delivery, replacement) with lower time-complexity. We categorize the related work based on the ML technique applied for prediction or clustering as these are the primary objective of ML application in caching enabled networks. The ML techniques within supervised learning, unsupervised learning, RL, NN, and TL are defined when first referenced. Interested readers can find a detailed overview of ML techniques in~\cite{Sun2019,Chen2019}. 
   
\subsection{Supervised Learning} 
Doan et al.~\cite{Doan2018} presented a caching scheme that predicts the popularity growth of published videos and estimates the popularity of new (unpublished) videos based on features extracted from published videos and their similarity with new videos. A backhaul link video traffic data minimization optimization problem that finds the optimal part of the video for caching is formulated and solved with linear programming. The optimization solution requires the popularity of both published and new videos as input. The spatio-temporal features of both published and new videos are extracted with a 3D convolutional neural network (CNN) with multiple stage pooling. CNN is a type of deep learning algorithm where convolution is applied in more than one layer of a multi-layer stacked network of convolutions and pooling. A G-dimensional vector represents features of a video and is mapped to the video categories. The clustering algorithm based on Jaccard and cosine distances are applied to extracted features to limit the curse of dimensionality. The number of features input to the clustering algorithm are restricted to reduce the complexity of the clustering algorithm. Support Vector Machine (SVM) is employed to classify each video into a video category. SVM is a type of unsupervised learning algorithm that finds a hyperplane in n-dimensional space to classify data points. A supervised ML model with published videos as training data is trained to predict the popularity of new videos while comparing similar video features. Expert advice is incorporated to update training data and evaluate the behavior of popularity predictor. The research does not take into account user mobility and the dynamic nature of the wireless network. However, the proposed work has the ability to predict the popularity of unpublished videos without historical information using supervised learning. The authors extend their work in~\cite{Doan2019} for socially-aware caching where social-awareness is measured as the frequency of D2D connections. 

Thar et al.~\cite{Thar2018} present a popularity based supervised and deep learning framework for proactive BS caching in MEC. The proposed cloud (master node) based framework predicts the video popularity using supervised and deep learning in two steps and pushes the learned cache decisions to the BSs (slave nodes). First, a data collector module collects data regarding request counts for video identifiers to predict future popularity and class of videos using supervised learning. Second, future request counts for video content are predicted using deep learning algorithms. An optimization framework is formulated to reduce the content access delay in future \textit{(t+1)} while making caching decisions at time \textit{t} based on cache capacity constraint. RNN and CNN models are evaluated for the training of popularity prediction and request count learning agents. The overall cache decision optimization problem is solved with deep NN which is composed of multiple levels of nonlinear operations and multiple hidden layers. The Movie Lens database is utilized as the data set. 

Authors in~\cite{Zhang2018} utilize learn-to-rank algorithm (supervised learning) and BS clustering (k-means) for caching in a small cell network scenario. The learn-to-rank does not directly use content popularity. Instead, the content is sorted based on its popularity rank. The objective of the study is CHR maximization and the optimization problem is formulated as integer programming whose solution is NP-hard if the content popularity is not known. The authors employ a learn-to-rank algorithm based on historical content requests. K-means clustering is applied to group SBSs based on historical request records for each file. K-means clustering algorithm partitions n data points into \textit{K} clusters such that each data point belongs to a cluster with the nearest mean. Afterward, a learn-to-rank algorithm is applied to predict the ranking of content popularity. Top one popularity is used to predict the probability of the number of requests for each file. 

The authors presented learning-based caching at SBS and UE in~\cite{Bastug2016,Bastug2014} to meet a required user satisfaction ratio. The authors considered proactive caching at SBS with the help of supervised learning. Each SBS learns a popularity matrix for its user with rows representing the users and columns representing content ratings. The missing matrix entries are inferred using collaborative filtering (recommendation systems). The proactive caching approach uses training and placement steps to offload traffic from backhaul links. In the first step, the popularity matrix is estimated while solving a least square minimization problem. The placement step is executed greedily until the caches are filled. In the second case study, the social relationship between users and their physical proximity is utilized for proactive caching at UE with D2D communications. The centrality measure that quantifies how well a user is connected is adopted for measuring the influence of social users. The authors utilize eigenvector centrality which is based on the largest eigenvalue of the adjacency matrix of the social network. Social influencers are grouped in communities with K-means clustering. The critical content of each community is stored in caches of social media influencers based on the premise that they are highly connected with the community for content distribution. The study is evaluated based on the dynamics of parameters such as the number of content requests, size of the cache, and zipf distribution parameter $\alpha$. The process of supervised learning in an edge network described in this  study consists of four steps. First, the user content rating and preferences are collected during peak activity hours. The content popularity is estimated with supervised learning algorithms. The most popular content with respect to a group of edge users is cached. User request are served from local cache in case of a cache hit.  

\subsection{Unsupervised Learning} 
Chang et al.~\cite{Chang2018} proposed a big data and ML-based framework for caching in edge networks. The authors provide two case studies to investigate smart caching in edge networks. In the first case study, unsupervised learning and deep learning are combined for the energy efficiency of data transmission optimization framework with constraints of user request latency. K-means clustering (unsupervised learning) is applied to group users in each cell based on user history and channel conditions. Selective scheduling is utilized to serve the end-user requests from the clustered groups. A DNN is trained to learn the behavior of optimal solutions based on input parameters (channel conditions and UE file requests) to reduce the time complexity of optimization. In the second case study, similarity learning is applied to investigate social ties between end-users where some users act as data receivers while others act as data transmitters (caches). A Kullback-Leibler-divergence-based metric is used to find similarity (common data interests) among UEs. This similarity (social tie) is then compared with D2D link quality. A one to one match between receiver and transmitter nodes is established based on the quality of social and physical ties.     

Liu et al.~\cite{Liu2020} proposed a proactive caching framework for next-generation cellular networks (HetNet) based on K-means clustering and content popularity prediction. The K-means clustering requires user location and content preferences as input to train the prediction model which can lead to privacy issues. It is advocated that traditional privacy-preserving algorithms based on symmetric encryption can lead to low user quality of experience due to their computational complexity. Therefore, the authors enhance the K-means algorithm as Privacy-preserving Federated k-means (PFK-means). A setup stage is utilized to share cryptographic keys among BSs and end-users for data authentication and signature. Subsequently, and update process performs the modified k-means algorithm in four phases namely, centroid initialization, mini-batch selection, mask key sharing, and centroid update. This results in cluster formation and cluster centroids are updated with encrypted data sharing when a user leaves or enters the cluster. The authors augment their framework with a light-weight privacy-preserving scheme that is based on secret sharing and federated learning. Federated learning conceptualizes usage of UEs to execute distributed ML model training instead of a centralized server. The UEs send ML model updates rather than data to the server, hence, preserving data privacy. 

\begin{figure*}
\centering
\includegraphics[width=13cm]{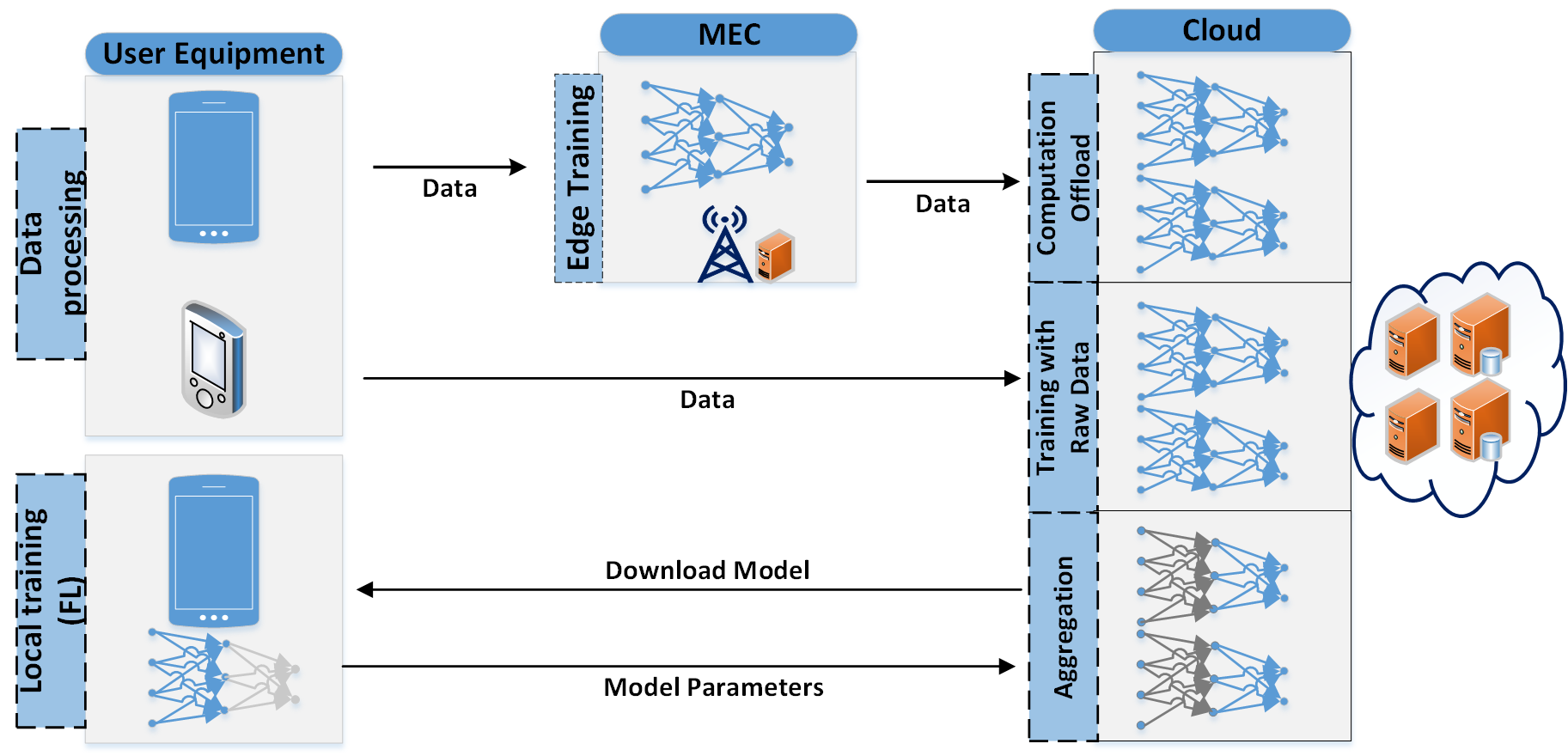} 
\caption {The Concept of Federated Learning for Privacy Preservation in Edge Networks}
\label{fig:fed}
\end{figure*}

The location of ML model can be centralized on a cloud server, collaborative between MEC and cloud server, or federated among multiple UEs and cloud. In the case of cloud computing, user data is sent to cloud servers to aid execution of ML models for training and prediction of caching related parameters. This approach incurs communication cost and may compromise user privacy during transmission. In the case of MEC, the users send their data to edge servers that execute ML models or further offload them to cloud servers. The collaborative MEC-cloud approach also leads to communication cost and privacy risk. Moreover, both MEC and cloud based approaches are unsuitable for applications that require constant training~\cite{Niknam2020}. Federated learning guarantees privacy as the data does not leave UE. Partial ML models are executed on the UE with local parameters such that the device is not computationally exhausted. The UEs collaboratively execute the ML model and send the output parameters of trained model to the FL server that aggregates the results. Figure~\ref{fig:fed} illustrates the difference between different learning approaches in edge networks (federated, MEC, cloud) to highlight the benefits of federated learning towards privacy-preserving mechanisms~\cite{Lim2020,Qian2019}.  
  
\subsection{Reinforcement Learning} 
Most of ML-based caching techniques in wireless networks apply RL to solve a sub-problem in their research method. The dynamic characteristics of wireless networks favor the application of RL techniques that can learn maximal rewards while interacting with the environment on trail and error basis~\cite{Zhu2018,Yu2019,Hu2018}. The learning process of an RL agent can be modeled as optimal control of Markov Decision Process (MDP)~\cite{He2017b} as described in the most of following listed studies. Delayed reward and trail and error are the two main features of RL. The first feature state that the RL agent looks does not consider immediate reward as optimal and looks for cumulative reward over a long time specified by a reward function. The second feature states that the RL agent finds a trade-off between exploration and exploitation. The agent exploits actions that have been proved to yield better rewards in the past while progressively exploring new actions that may yield better rewards~\cite{Yu2019,Zhu2018}.  

A BS based distributed edge caching and D2D content delivery framework was presented in~\cite{Wang2017a}. The focus of the article in on edge cache replacement technique. An MDP optimization problem is formulated that minimizes cache replacement transmission cost based on variables such as content popularity, cellular serving ratio, and communication cost of cache replacement from one BS to another. A Q-learning algorithm is adopted to solve the optimization problem of transmission cost minimization while identifying cache replacement at each BS. Q-learning is a type of RL algorithm that interacts with its environment to learn Q-values based on which the learning agent takes action. The cellular serving ratio for the static network is calculated from the maximum weighted independent sets of a conflict graph. In the dynamic network scenario, the cellular serving ratio is calculated using probabilistic methods. The cellular serving ratio helps determine the caching capability of users with D2D communication. As the cache replacement process is distributed at each BS, each BS calculates its replacement reward based on the transmission cost of cache replacement. The Q-learning algorithm measures content placement as observed state and cache replacement as action. 

Researchers~\cite{Xiang2019,Xiang2020} consider a DRL approach for content caching and Fog Radio Access Network (F-RAN) slicing in vehicular network scenarios. F-RAN slicing is considered to identify user hotspots in the network. The variance in wireless network conditions and content popularity due to spatio-temporal nature of the content and wireless networks increases problem complexity. The user content requests are modeled as zipf distribution while the wireless channel is modeled as a finite-state Markov channel. The APs are connected to the cloud server which executes the optimization solutions and pushes content to appropriate access point (AP) for caching. The server executes DRL based optimization solution which is formulated based on constraints of fronthaul capacity and AP capability to determine user device mode and content caching. DRL, also known as deep Q Network (DQN), utilizes NN to approximate the Q-values of the Q-learning algorithm. The research does not consider the mobility of users while formulating the problem in vehicular networks. 

Authors in~\cite{Liu2019} proposed a two-step approach for content recommendation and caching with an overall objective to minimize MNO cost where each step utilizes a separate RL agent. The first step of their approach is the content recommendation which works without prior knowledge of user preference. If the mobile user does not accept the recommendation, the experience is used to reinforce the recommendation model. A central processor connected to multiple BSs collects information on user requests and channel conditions. The central processor collects this data in a database that is used to train the recommendation model based on actor-critic network. User stickiness ensures that the mobile user is more likely to accept recommendation if preferred content is recommended. The well-learned recommendation agent helps reduce the state space of the second RL agent that learns to cache and avoids the curse of dimensionality. The recommended content is proactively and opportunistically cached to the mobile device if channel conditions are suitable. The user mobility pattern is modeled as a Markov Decision Process. The objective of the caching agent is to reduce transmission delay and improve user QoE. The overall optimization problem is solved by double deep Q-network (DDQN) with dueling architecture. The proposed approach is compared with other policy-based RL approaches in a simulated environment to debate the optimization of MNO's revenue.   

\begin{figure}
\centering
\includegraphics[width=8.5cm]{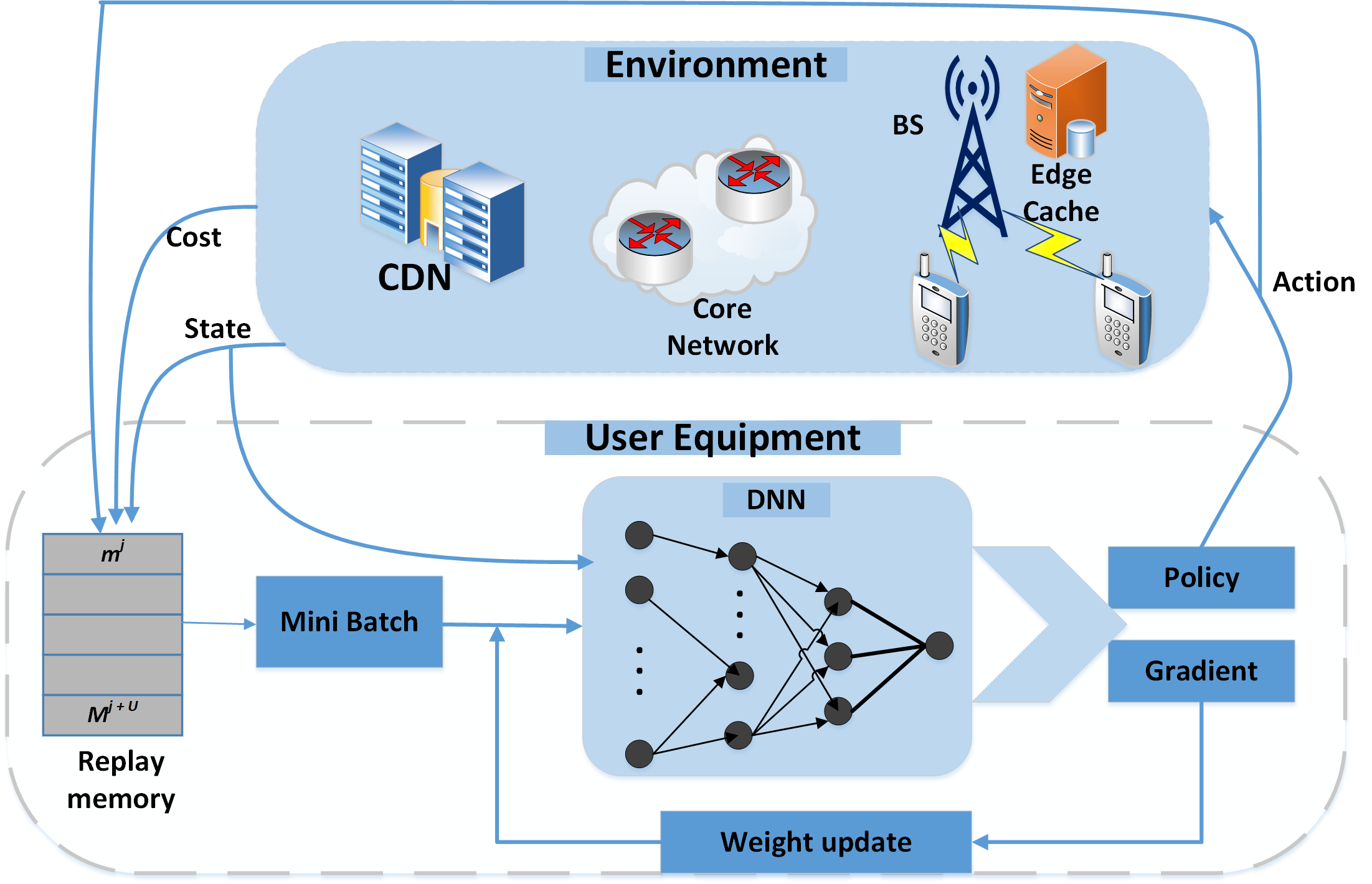} 
\caption {The Concept of DQN Algorithm in Edge Networks~\cite{Chen2018a}}
\label{fig:dqn}
\end{figure}

Researchers~\cite{Jiang2019} proposed a distributed multi-tier (BS, AP, UE) content caching scheme based on deep Q-learning to increase the efficiency of F-RAN. The scheme first predicts the content popularity from historical user preferences. Instead of topic similarity and community detection, a topic modeling approach named probabilistic latent semantic analysis (pLSA) is applied to predict individual user preferences. pLSA algorithm requires historical user requests for cloud content and topic variables that represent the relationship between content and topics. Expectation maximization (EM) algorithm is used to calculate the probability that a user requests a specific cached content. Collective content popularity for all users is predicted based on current caching status and transmission distance between BSs. The DQN algorithm consists of two individual CNN. One of the CNN (MainNet) predicts current Q-values while the other CNN (TargetNet) predicts Q-values for a certain time period for the case of content caching decision. If the content is not found in the cache, an update process is initiated based on new user preferences, subsequently, updated content popularity. A generic DQN consists of a replay memory of finite space to save transitions regarding system state and cost at the end of each epoch. A mini-batch of randomly selected transitions are used to train a DNN to minimize the loss function and update caching policy. The DNN weights are updated to the next epoch and transition. The proposed DQN is depicted in Figure~\ref{fig:dqn}.

\begin{figure}
\centering
\includegraphics[height=6cm]{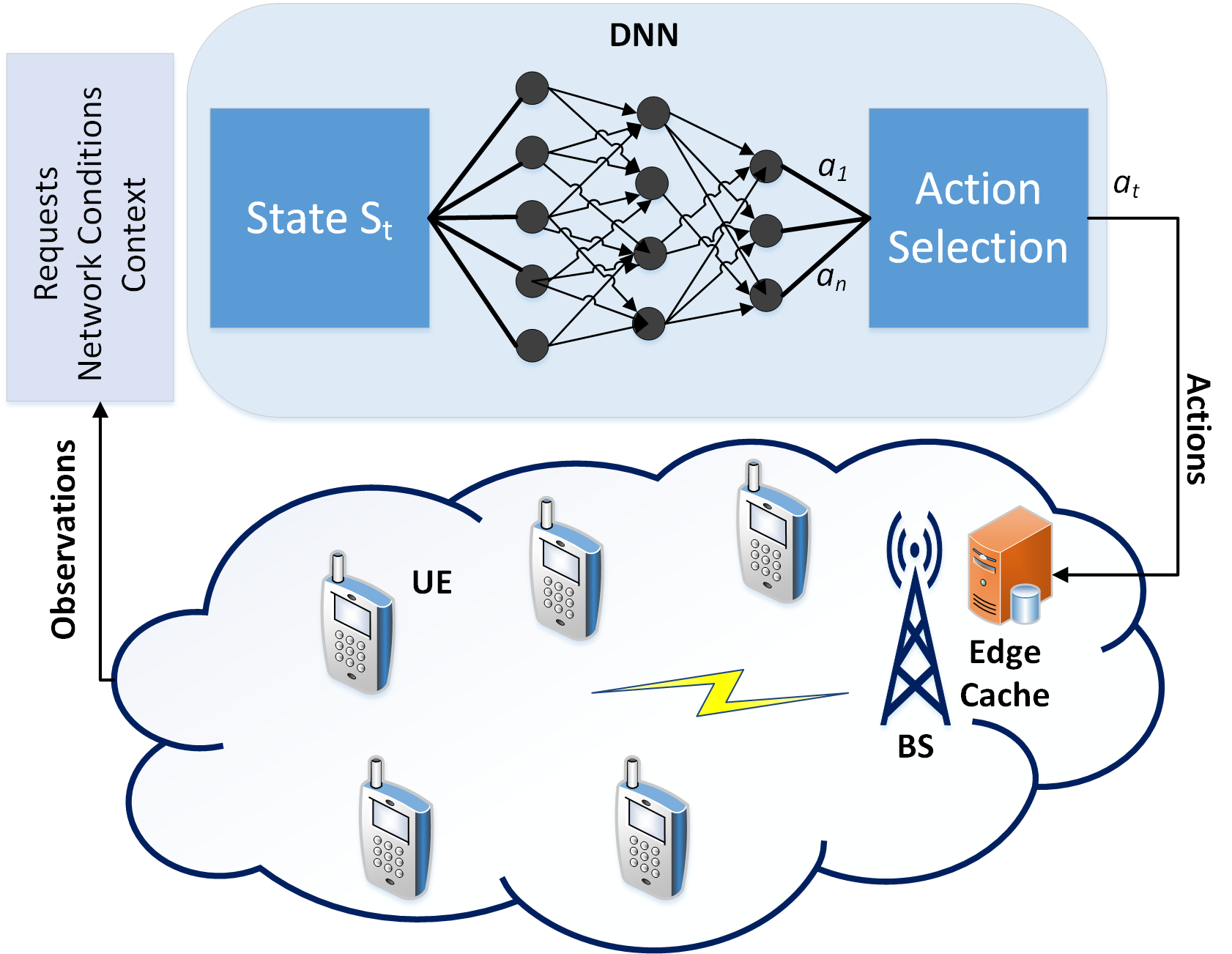} 
\caption {Reinforcement Learning for Edge Caching~\cite{Zhu2018}}
\label{fig:rl}
\end{figure}

Jiang et al.~\cite{Jiang2019a} present a multi-agent-based RL framework for D2D caching networks with the objective of download latency minimization. The D2D cache optimization problem is formulated as a multi-agent multi-armed bandit (RL) where UEs act as both joint and independent learners, content is considered as arms and caching as actions. The joint and independent learners learn from content demand history in the absence of knowledge of content popularity distribution. In the single-agent MAB model, an agent learns the reward of his sequential actions with the aim to maximize overall reward. In multi-agent MAB mode, there exists a trade-off between current best action and gathering information for achieving a higher reward for future actions. In the independent learning scenario (single agent), UEs employ Q-learning to coordinate the caching decisions in a large action space which is reduced by modified combinatorial upper confidence bound algorithm. In the joint learning scenario (multi-agent), UEs learn Q-values of their caching decisions along with the actions of other UEs. The Q-values are also considered for cache replacement decisions in the next optimization interval. Simulations show the superior performance of joint learning with respect to independent learning in terms of CHR and download latency. Interested readers can refer to a short review of DRL techniques for edge caching in~\cite{Zhu2018}. A RL observes network conditions, user requests, and context as a state of the system at time \textit{t}. The system state is input to train the DNN and predict actions that will lead to maximum benefit in caching decisions. Figure~\ref{fig:rl} illustrates a generic RL based edge caching framework. 

\subsection{Neural Networks}
\begin{figure}
\centering
\includegraphics[height=7cm]{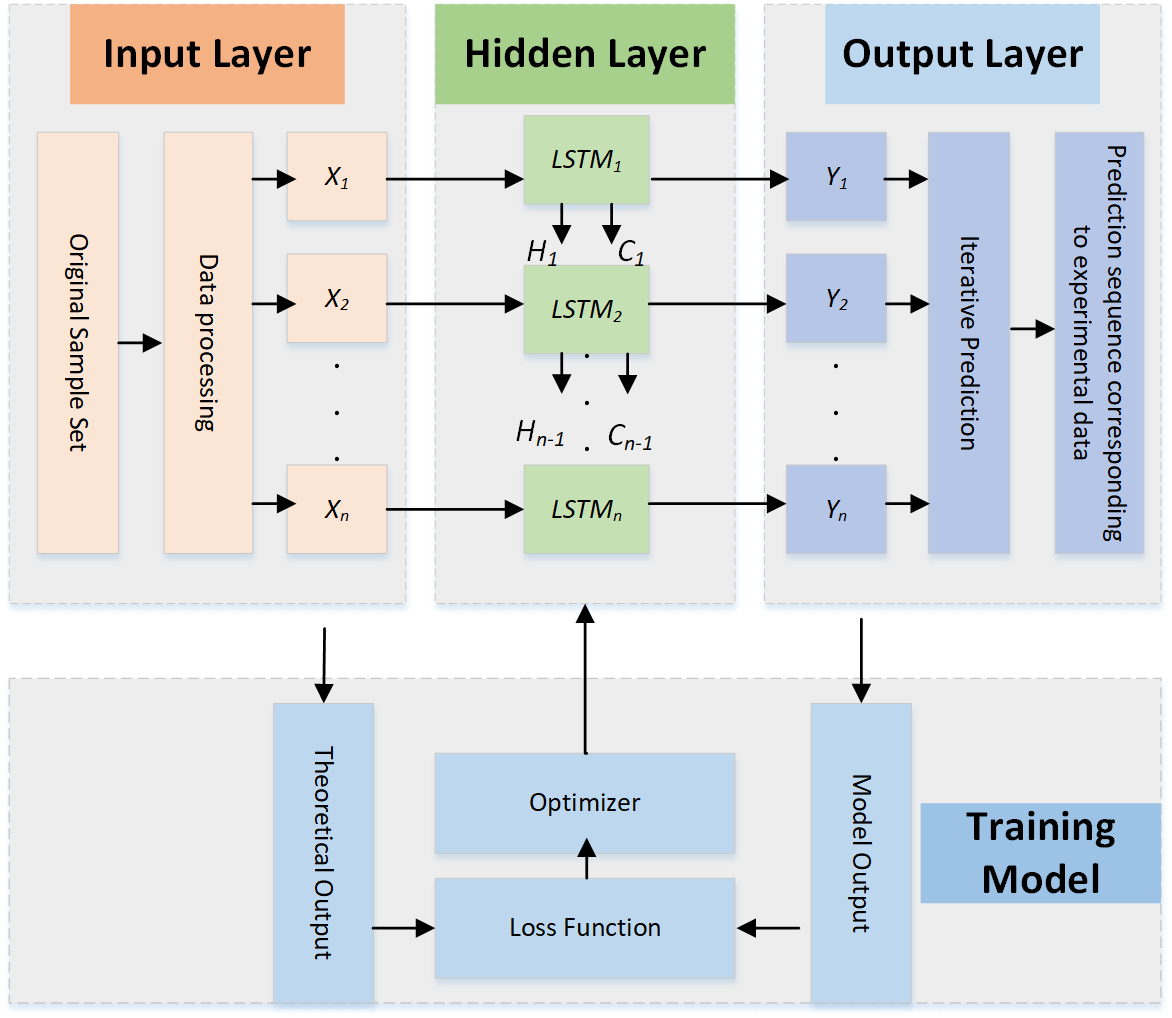} 
\caption {RNN based long-short term memory (LSTM) model for popularity prediction~\cite{Li2019}}
\label{fig:nn}
\end{figure}

The authors of~\cite{Li2019} propose a cache placement and delivery mechanism for D2D networks with the goal of traffic and delay minimization. The proposed methodology answers what and where to cache while predicting user mobility and content popularity. A recurrent neural network (RNN) based echo state network (ESN) model is utilized to predict the next location of the user based on his long-term historical information in a D2D network. RNN is a type of feed-forward NN with internal memory and applies the same function to each input. ESN is a variant of RNN with a sparsely connected hidden layer. Further, an RNN based long-short term memory (LSTM) model is applied to predict the user content popularity based on multiple features such as age, gender, time, and location of content request. LSTM is a modified version of RNN which remembers past data in memory and resolves the RNN problem of vanishing gradient. Better user mobility and content popularity prediction leads to better content placement, hence, higher cache hit ratios. Afterward, the D2D content delivery decision (establish D2D link with which neighbor for cached content) is based on the DRL algorithm which optimizes the data transmission energy and content access delay while taking into account variables like channel state and distance between UE. A reward function based on RL is proposed to learn the reward of caching actions for future caching updates. The LSTM (RNN) based content popularity prediction model is depicted in Figrue~\ref{fig:nn}. It consists of four modules, i.e., input layer, hidden layer, output layer, and training network. The input layer pre-processes the data set so that it meets the requirements of the standard input. The hidden layer employs a RNN constructed by multiple LSTM cells. The network is trained by the back propagation and the loss function is optimized by the optimizer. The predicted output values can be iteratively updated to the output layer.  

Researchers~\cite{Tanzil2017} presented a video content caching scheme for cellular networks to improve user QoE and reduce backhaul traffic. The scheme takes input user preferences and cellular network properties such as bandwidth. The proposed technique formulates an extreme learning machine (ELM) to predict the future popularity of content based on features extracted from a stochastic approximation algorithm. ELM is a type of feed-forward NN with hidden layers whose parameters are not tuned. The ELM algorithm is utilized in a scenario where the YouTube video features are noisy. Video features in the context of human perception are broadly classified into four categories, i.e., title, keyword, channel, and thumbnail. A total of 54 features are selected for each video using human perception models that take sentiment, subjectivity, video brightness, and contrast as input among other parameters. Stochastic Perturbation Simultaneous Approximation (SPSA) is employed to select a sample number of neurons for the ELM such that its computational complexity is reduced for online popularity prediction while the predictive performance is not compromised. A mixed-integer linear programming optimization is utilized to decide the cache placement (BS station) based on predicted popularity and cellular network properties. Cache replacement utilizes an improved version of segmented least recently used (SLRU) algorithm. The study considers global popularity features extracted from YouTube. On the contrary, video preferences for a network edge-specific community and the mobility of users need to be considered for edge networks. the authors extend their work with risk-neutral and risk-averse caching algorithms for forecasting of content distribution function~\cite{Hoiles2018}. 

Authors in~\cite{Lei2018} proposed a proactive caching strategy based on deep NN and unsupervised learning in 5G networks. Firstly, SDN/NFV technologies are implemented in network core to gather data and utilize ML functions, specifically a Stacked Sparse AutoEncoder (SSAE) at distributed network elements. An autoencoder is a type of NN that learns data coding in an unsupervised manner. A virtual server receives content statistics from SDN based network resources to predicts content popularity using deep learning. The caching strategy is learned by the SDN controller and network nodes are informed of cache placement. An NN based feature extraction method (SSAE) is used to predict the distribution of user content requests. SSAE employs two stacked auto encoding layers for feature extraction with each stage trained by an unsupervised learning algorithm. The content popularity is utilized in learning of caching strategy where optimal solutions are reached by Deep NN to reduce time-complexity.  

Chen et al.~\cite{Chen2017} study the problem of proactive caching with the help of unmanned aerial vehicles (UAV) in the CRAN scenario while aiming to maximize the user QoE. The UAV and a virtual pool of BBUs cache content to service ground and terrestrial devices with the help of the mmWave frequency spectrum to reduce blocking effects. The content can be delivered by CRAN based virtual BBU pool, UAV, or neighbor devices with D2D communication. Content popularity and user mobility prediction are learned at virtual BBUs with the help of ESN based on human-centric parameters, such as user locations, requested contents, gender, and job. A conceptor-based ESN is proposed for popularity distribution and mobility prediction to solve the cache optimization problem that maximizes user QoE and minimizes UAV transmit power. The prediction of user content requests and mobility derives user-UAV associations. The user-UAV associations lead to optimal positioning of UAV and cached content where nearby users are grouped with k-means clustering such that one cluster is served by one UAV. The mobility model assumes constant speed of the user, with user location monitored after every hour. The UAV remains static while content is transmitted to a user. The ESN (RNN) model employs conceptors to enable prediction of multiple user mobility and content popularity distributions over different time frames i.e., days or weeks.

\subsection{Transfer Learning}
Authors in~\cite{Bharath2016} utilize a TL approach to transfer the popularity of content from social networks to a HetNet in order to optimize CHR. The SBS placed caching strategy assumes user distribution as poison point processes. The unknown popularity profile of cached content is estimated using instantaneous user requests in a time interval at an SBS. Predicting random poison distributions for request arrival is time-consuming with variable network parameters. TL helps improve the popularity prediction while transferring knowledge of popular content from sample social network content. Users connected to SBS are analyzed for social community behavior. The social community acts as the source domain whose knowledge is transferred to the target domain (user request pattern). 

\begin{figure}
\centering
\includegraphics[height=6.2cm]{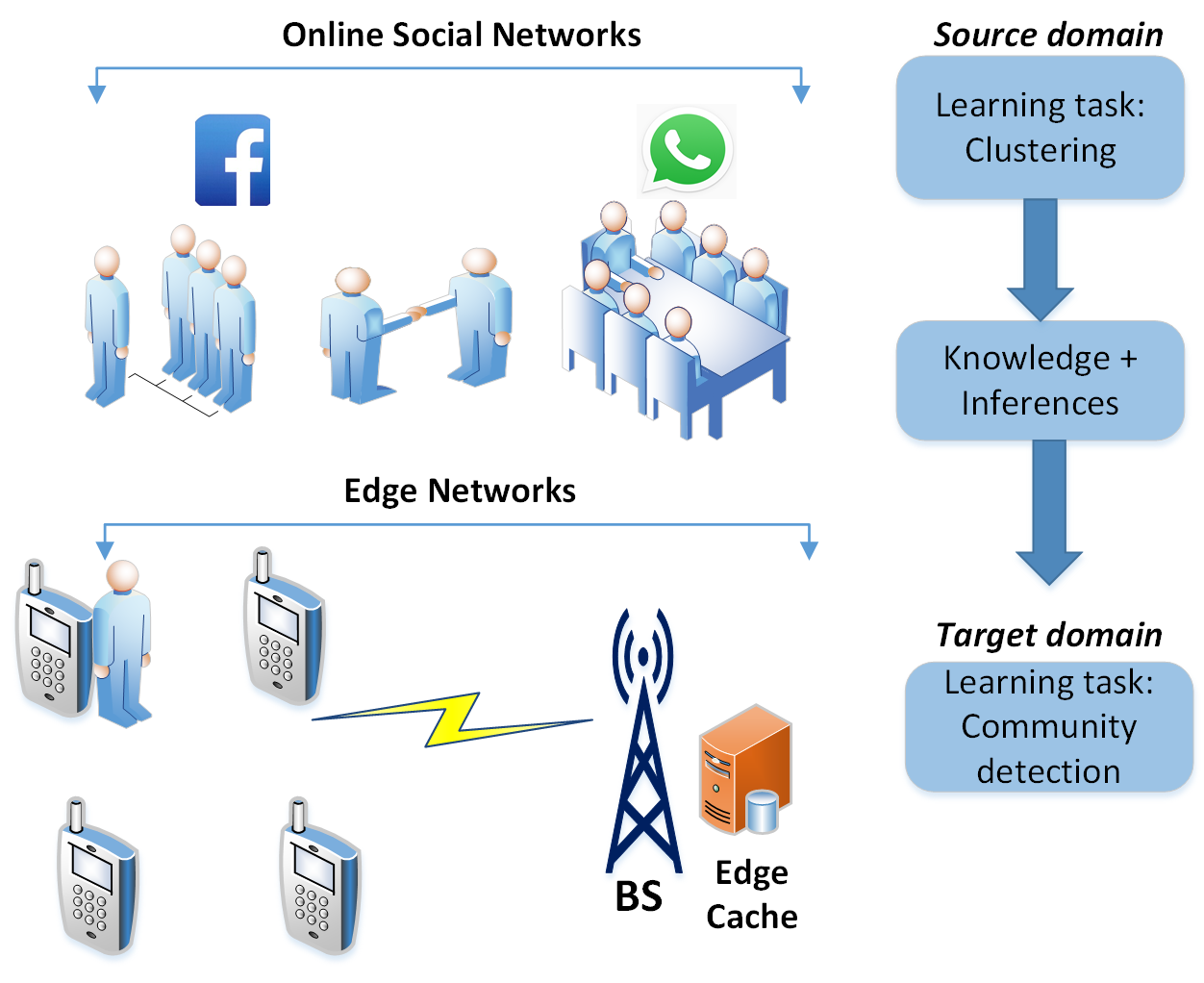} 
\caption {Transfer Learning for Edge Caching}
\label{fig:tl}
\end{figure}

Authors in~\cite{Hou2018} present an unsupervised learning and TL scheme for proactive and distributed caching in MEC without prior knowledge of content popularity. Instead of learning a task from scratch, TL agents transfer knowledge from relative tasks with fewer training data to speed up the learning process. The learning-based caching scheme utilizes MEC and distributed BSs as cache locations to reduce data transmission costs. The caching optimization problem formulated as an integer programming problem is proved as NP-hard. A proximate solution is provided by a greedy algorithm that optimizes a parameter named as maximum transmission cost reduction. Two sub-problems of content classification and popularity prediction are solved by a 2 step learning strategy aided by unsupervised learning and TL. K-means clustering algorithm takes historical access features of the content during a time interval as input in the form of a 2D vector. After classification of content into K classes, TL is employed to predict class popularity. The higher correlation between historical access and cached content, the better is the performance of the TL algorithm. The popular content is input to the cache optimization problem to decide distributed caching places. The process of transferring knowledge from online social networks to edge networks for content caching is illustrated in Figure~\ref{fig:tl}. In general, social networks can be explored for social ties between edge user and the topics that interest users. This knowledge can be transferred to the clustering algorithms that can group users or their data into communities with respect to an edge network.    

\subsection{Other Machine Learning Techniques}
Bommaraveni et al.~\cite{Bommaraveni2019} present an active query learning (semi-supervised) approach for accurate content request prediction in 5G Edge networks. The basic premise of the active learning algorithm is to present users with recommendations to learn their preferences. SBS with non-overlapping coverage and attached edge servers are considered. User requests are modeled as a demand matrix where a 0 or 1 entry is marked against each available content. The missing entries of the demand matrix are estimated by active learning-based Query-by-committee (QBC) matrix completion algorithm. The basic intuition of QBC algorithm is to form a committee of matrix completion algorithms that aim to minimize the version space with the aim to find most informative missing matrix entry that accelerates the learning rate. The SBS cache is divided into two portions. One caches the content predicted by the demand matrix for CHR exploitation while another caches content based on an uncertainty matrix for further user behavior exploration. 

The article by Chuan et al.~\cite{Chuan2019} investigates an ML-based content popularity regeneration framework in wireless networks. The BSs cache popular content based on the common interests of mobile users within their service range. The authors prove that the popularity prediction distribution of mobile users' common interest is based on multi-dimensional Dirichlet distribution. The high computational complexity of popularity distribution requires a sampling-based solution which is derived from a learning-based Gibbs sampling model. The learning model for popularity prediction is based on historical content delivery data set. Based on the assumption that content popularity is derived and learned, content caching optimization is converted into a decision-making problem. The system model lacks details of user mobility, a content delivery technique which are necessary for cellular networks. 

\begin{table*}
\centering
\caption{Comparison of ML based Edge Caching Techniques}\label{tab:compmcc}
\resizebox{0.84\textwidth}{!}{
\begin{tabular}{|p{1.1cm}|p{1.8cm}|p{2.25cm}|p{2.6cm}|p{2.2cm}|p{1.2cm}|p{1.15cm}|p{1.2cm}|p{1.25cm}|}\hline
\textbf{Study} & \textbf{ML method} & \textbf{ML Objective} & \textbf{Features and patterns} &  \textbf{Caching Objective} & \textbf{Caching Policy} & \textbf{Location} & \textbf{\makecell{Replace-\\ment}} & \textbf{Network scenario} \\ \hline 
~\cite{Doan2018} & Supervised, CNN &  Popularity prediction, Feature extraction & popularity of published content & Increase offload ratio & Proactive & MBS & NA & Macro-cellular \\[0.1cm] 
~\cite{Thar2018} & Supervised, DNN  & Popularity prediction, cache optimization & request count based popularity & decrease access latency & Proactive & BS & NA & MEC  \\
~\cite{Zhang2018} & Supervised, Unsupervised & Popularity ranking, BS clustering & user preference based popularity  & Increase CHR & Reactive & SBS & NA  & Small cell network \\
~\cite{Bastug2016} & Supervised learning  & popularity prediction & request count and rating based popularity & increase offload traffic, user satisfaction & Proactive & UE, BS & NA & Small cell network \\[0.1cm]\hline 

~\cite{Chang2018} case study 1 & Unsupervised, DNN  & user clustering, cache optimization & request count and channel conditions  & Decrease latency, EE data transmission & Proactive & MBS & NA & Macro-cellular, D2D  \\[0.1cm]
~\cite{Chang2018} case study 2 & Unsupervised  & similarity learning & user preference & Maximize D2D data rate & Proactive & UE & NA & D2D \\[0.1cm]
~\cite{Liu2020} & Unsupervised (K-means)  & user clustering & user preference based popularity and location  & Minimize communication overhead & Proactive & MBS, SBS  & NA & HetNet  \\[0.1cm] \hline

~\cite{Wang2017a} & RL (Q-learning)  & Cache replacement  & request count based popularity, poison point based user distribution & cache replacement transmission cost & Proactive & UE, BS & Q-learning & Macro-cellular, D2D \\[0.1cm] 
~\cite{Xiang2019} &  RL (Deep)  & RAN slicing & request count based popularity, markov variable based channel coefficients & Maximize CHR, transmit rate  & Reactive & UE, V2I & NA & F-RAN  \\[0.1cm]
~\cite{Liu2019} & RL  & Recommendation, caching optimization & user request and channel condition & Reduce MNO cost & Proactive & MBS, UE & NA & Macro-cellular \\[0.1cm]
~\cite{Jiang2019} & RL (DQN), CNN & caching optimization, predict Q-values & user preference based popularity & Maximize CHR & Proactive & UE, BS, AP & User-preference  & F-RAN\\[0.1cm]
~\cite{Jiang2019a} & RL (MAB, Q-learning)  & caching optimization, predict Q-values & zipf based popularity distribution and access & Download latency optimization & Proactive & UE, BS, AP & Q-learning based & D2D network \\[0.1cm]
\hline

~\cite{Li2019} & RNN (ESN, LSTM), RL & mobility and popularity prediction, problem optimization & user context based popularity, history based locality & Improve CHR, reduce transmission delay and energy & Proactive & BS, UE & FIFO &  D2D network \\[0.1cm]
~\cite{Tanzil2017} & NN (ELM) & Popularity prediction & content feature based popularity & Increase QoE, decrease network delay  & Reactive & MBS & Segmented LRU & Macro-cellular\\[0.1cm] 
~\cite{Lei2018} & DNN, AE  &  feature extraction, problem optimization  & historical content request information & Increase CHR & Proactive & Core, distributed BSs  & NA & Distributed core \\[0.1cm] 
~\cite{Chen2017} & RNN (ESN)  & mobility and popularity prediction & user context based access and location distribution & increase QoE, decrease transmit power & Proactive & UAV, UE, BBU & NA & CRAN \\[0.1cm] \hline
~\cite{Bharath2016} & TL & Transfer popularity from social network  & Poisson point distribution based content request & increase CHR & Proactive & UE, SBS & NA & HetNet \\[0.1cm]
~\cite{Hou2018} & Unsupervised (K-means), TL & content clustering & user access based popularity  & Minimize data transmission cost & Proactive & MEC, BS  & NA & MEC \\\hline 
~\cite{Bommaraveni2019} & Active query learning  & popularity prediction & user access based popularity & increase CHR, decrease backhaul load & Proactive & UE, SBS & NA & Edge network \\[0.1cm]
~\cite{Chuan2019} & Supervised & Feature extraction & user interest based popularity & Increase CHR & Proactive & BS & NA & Macro-cellular \\\hline

\end{tabular}}
\end{table*}

\section{Comparison of ML based edge caching techniques}\label{sec:comparison}
This section provides comparison of the listed studies. The comparison is summarized in Table~\ref{tab:compmcc}. 

In the context of edge network caching, ML techniques enable edge devices to actively and intelligently monitor their environment by learning and predicting various network features, such as, channel dynamics, traffic patterns, and content requests to proactively take actions that maximize a predefined objective such as, QoS and CHR~\cite{Chen2019}. It can be observed that most of the state-of-the-art techniques apply more than one ML techniques to solve research problems related to optimal edge caching. Supervised learning techniques are mostly applied for content popularity prediction where initial data is available for model training. Supervised learning algorithms are suitable when labeled data is available in the form of historical network state information~\cite{Doan2018}. Unsupervised techniques are applied for clustering (K-means) users, content, and BSs based on their location or access features~\cite{Thar2018,Chang2018}. Unsupervised learning techniques explore unlabeled data such as user geo-location to infer some structure or pattern. RL algorithms are applied to solve the overall problem of cache optimization (where and what to cache) given the network states, actions, and cost~\cite{Jiang2019}. In edge cached networks, caching decisions involve multiple dynamic inputs including mobility, popularity, and access distribution that vary with time, As a result, the application of a RL agent is well suited to edge cache optimization as it interacts with the environment on trail and error basis to to optimize an objective. The RL agent learns and trains on the data collected from the implementation of RL. NNs learn from complex and imprecise data and are applied to forecast the mobility of users and the popularity of the content~\cite{Li2019}. NNs can predict user and network characteristics to help solve complex cache optimization problems and design optimal caching strategy. The data in wireless edge networks is continuous and time-dependent. RNNs (LSTM and ESN) are suitable for dealing with time-dependent data~\cite{Chen2019,Li2019}.

The ML techniques applied for caching in edge networks mostly employ RL and NN. The main reason behind this is the unavailability of data for the prediction of content popularity and user mobility. RL and NN techniques can learn content popularity with little or no data~\cite{Jiang2019}. Therefore, NN and RL techniques have larger applications in edge caching networks~\cite{Sun2019,Ye2019}. Moreover, RL, NN, and deep learning are more sophisticated learning techniques than supervised and unsupervised learning. Hence, they are more suited for the dynamics of edge networks. Where data is available, supervised and unsupervised learning are utilized~\cite{Zhang2018}. Among supervised learning SVM~\cite{Doan2018}, in unsupervised learning K-means clustering~\cite{Hou2018,Liu2020}, in RL Q-learning~\cite{Wang2017a} and Deep RL~\cite{Xiang2019}, and in NN Deep NN~\cite{Li2019} are widely utilized to address different issues within optimal caching. SDN, NFV, and C-RAN technologies can be applied in network devices to maintain user content request history which can be inspected for future content popularity prediction~\cite{Lei2018}. Monitoring user request history also leads to privacy issues that have been seldom addressed in recent research~\cite{Liu2020}. In order to gain user faith for participation in data set collection activities, light-weight privacy-preserving, authentication, and trust protocols are necessary for MEC~\cite{Khurshid2019,Li2018a}.   

Supervised learning and NN are mostly applied for popularity and mobility prediction~\cite{Chen2017,Doan2018}. Unsupervised learning techniques are mostly applied for clustering~\cite{Liu2020}. Feature extraction from content is done with the help of NN~\cite{Lei2018} or supervised learning~\cite{Chuan2019}. RL and NN algorithms are used for cache decision optimization~\cite{Liu2019,Li2019}. Q-learning is appropriately employed for cache replacement by learning Q-values/rewards of caching actions~\cite{Wang2017a}. TL is applied to transfer knowledge of content popularity from social networks to mobile networks~\cite{Bharath2016}.  

The caching objective of the majority of cited work is increasing CHR~\cite{Zhang2018,Lei2018}. All other caching objectives are directly related to CHR fluctuations. Content access latency, user QoE, data offload ratio, and MNO's operational cost is immediately effected when the content is available in-network and not accessed through CDNs and backhaul links. Hence, the single objective of CHR maximization is often sufficient to address caching optimization problems. The research works that employ D2D communications for delivery strategy often have the objective to reduce data transmission costs (energy efficiency) at UE~\cite{Li2019}. As UE cache and share data with other UEs within the network with little or no incentives, the energy efficiency of the sender is desired. The objective of lower data transmission costs is synonymous with minimal communication overhead~\cite{Liu2020}, maximum data transmission rate, and energy-efficient data transmissions~\cite{Chang2018}. Most of the state-of-the-art works in caching are proactive rather than reactive to further offload traffic from backhaul and core networks~\cite{Doan2018}. The main reason behind proactive caching in ML-based techniques is that the learning process can start without the user initiating a content request. On the other hand, statistical techniques for caching, e.g., most frequently accessed, are mostly reactive to user behavior.  

Cache delivery networks and location options are diversely employed in the listed literature work~\cite{Doan2018,Thar2018}. This comparison shows that the studies do not focus on one network scenario or one cache location. Each of the choices about the cache delivery network and location has its own pros and cons. Placing cache at UE further lowers access latencies and offloads traffic from both backhaul and fronthaul networks. Therefore, caching techniques that utilize UEs offer better user QoE than BS caching based techniques~\cite{Li2019}. However, UE caching puts the burden of energy and storage resources of handheld devices without a standard incentive policy~\cite{Jiang2019}. MEC, F-RAN, and C-RAN infrastructure based studies assume knowledge of content request history from in-network device statistics~\cite{Thar2018,Jiang2019}. Therefore, MEC /F-RAN / C-RAN based studies have the advantage of user content history over macro-cellular, D2D, and HetNet based network architectures. Moreover, SDN/NFV based techniques are necessary to gain user statistics from n network devices~\cite{Lei2018}. Furthermore, MEC/F-RAN/C-RAN based techniques have lesser complexity in user mobility prediction as their coverage is wider than macro-cellular D2D networks~\cite{Liu2019}. An innovative idea for cache location was presented in~\cite{Chen2017} where UAV where utilized to cache content within the network to support low access latencies in user dense areas. However, the proposal will incur a higher operational cost as compared to standard network cache locations. Among the listed works only one considers the joint RAN slicing and caching in edge networks~\cite{Xiang2019} and one considers user privacy as part of the edge caching solution~\cite{Liu2020}. 

We have not included the comparison of delivery strategy in Table~\ref{tab:compmcc} as all studies utilized implicit unicast or explicit D2D delivery in edge caching. Multicast and CoMP based delivery strategies have not been discussed which leaves room for considerable optimization in existing studies~\cite{Bilal2019,Said2018}.   Cache replacement techniques have not been focused on in most of the cited work. This indicates that a cache optimization problem solved for time \textit{t} can be updated for time \textit{t+1} for cache replacement decision. However, the repetition and recursive call of the cache optimization problem can incur significant complexity. Therefore, a simpler cache replacement decision is desired. Wang et al.~\cite{Wang2017} prove that a simple Q-learning model can be utilized to learn cache rewards from the decisions based on time \textit{t} and update cache replacement decision at time \textit{t+1}. Some of the studies apply statistical caching approaches that have limited benefit in a dynamic network environment~\cite{Tanzil2017,Li2019}. Cache replacement is studied independently as an MDP problem with a Q-learning solution in~\cite{Gu2014}.

Few references utilize YouTube~\cite{Tanzil2017} and MovieLens~\cite{Adeel2019,Thar2018} data-sets for performance evaluation of proposed popularity prediction techniques. Therefore, most of the studies use statistical models such as zipf and PPP to model input variables to cache optimization problem. Four network dynamics not detailed in~~\ref{tab:compmcc} need to be debated among the listed work. These are \textbf{(a)} content popularity distribution, \textbf{(b)} content request distribution, \textbf{(c)} user mobility model, and \textbf{(d)} network state model. The benchmark data-sets or models for content popularity, user request, mobility, and network state prediction are not available for edge networks resulting in non-baseline performance evaluations.

Most of the works consider the content popularity to be static following the zipf distribution model~\cite{Liu2019,Xiang2019,Bharath2016,Wang2017a,Chen2018b,Hou2018}. The zipf model is based on power law distribution and utilized for web content popularity~\cite{Tatar2014}. In general, the content popularity changes in web caches are very slow as compared to dynamic edge networks with user mobility characteristics~\cite{Liu2016}. Therefore, the application of static zipf model to content popularity distribution in edge networks is not suited. Some researchers have adopted dynamic content popularity distribution models based on user preferences and topological relationships as in~\cite{Jiang2019,Jiang2019a}, as Markov process in~\cite{Sadeghi2017}, and ML-based popularity prediction in~\cite{Tanzil2017}. 

User content requests are predominantly modeled as Poisson point distributions~\cite{Li2019,Wang2017a,Hu2018,Jiang2019a,Hou2018}. Researchers~\cite{Tanzil2017} further utilize ELM for to predict content popularity based on \textbf{(a)} content features built from human perception model and \textbf{(b)} user requests modeled as Poisson distribution. Some researchers model user requests as zipf distribution~\cite{Bommaraveni2019,Bastug2016} or Markov process~\cite{Sadeghi2019}. 

Some of the researchers ignore user mobility~\cite{Doan2019} while others model it as Poisson point process (PPP)~\cite{Hu2018,Jiang2019a}, random walk model~\cite{Alfano2014,Malik2020,Chen2018b}, or as Markov process~\cite{Liu2019}. Bastug et al.~\cite{Bastug2016} consider the location of Cache enabled BSs distributed as PPP. Several works utilize NNs to predict the mobility of users in edge networks. Li et al.~\cite{Li2019} utilized RNN based ESN model to predict user mobility and LSTM based model to predict content popularity. The ESN provides non-linear system forecasting with temporal inputs. The ESN model takes location history of a user over a period of time as input and initially modeled as PPP to predict user history at \textit{t+1}. The output of ESN model for user location and user context (gender, age, occupation, time, etc) are input to the LSTM for content popularity prediction. Similarly, Chen et al.~\cite{Chen2017} utilize modified ESN to predict the user content request distributions and mobility patterns. User context information (time, week, gender, occupation, age, location, and device type, etc.) is input to the ESN to get output of content request and mobility patters. The initial user mobility is adopted from pedestrian mobility patterns according to which a user visits places of interest at a specific time of the day. Researchers~\cite{Yin2018a} use user location history as input to ESN to predict mobility patterns and user context as input to another ESN to predict content popularity. 

The dynamic wireless channel state in terms of Channel State Information (CSI) is modeled in few of the studies~\cite{Chuan2019,Li2019,Wang2019a}. Researchers~\cite{Xiang2019,Wang2019a,Hu2018} models the wireless channel as finite-state Markov channel. Chen et al.~\cite{Chen2017} model the mmWave propagation channel between the UAVs and users using the standard log-normal shadowing model. Researchers~\cite{Li2019,Yin2018a} model wireless channel with dynamic path loss, channel attenuation index, and Gaussian noise. Other studies assume channel conditions to be invariant.

Many studies model some parameters of the optimization problem as MDP. For example, reference~\cite{Sadeghi2017} model user content request as Markov process with unknown transition probabilities. References~\cite{Gu2014,Wang2017a,Wang2019a} model the cache replacement problem as MDP. Reference~\cite{Liu2019} mobility patterns of users are characterized as Markov process. Reference~\cite{Xiang2019} models the wireless channel as finite-state Markov channel while reference~\cite{He2018} models cache state as Markov model.
\section{Research Challenges and Future Directions}\label{sec:cfd}
The listed state-of-the-art research works have provided the foundations of ML application in edge caching while addressing issues of popularity prediction~\cite{Trzcinski2017,He2017}, mobility prediction~\cite{Chen2017}, social awareness~\cite{Wu2020,Shan2019}, community detection~\cite{ElBarawy2014,Nguyen2019}, cache decision optimization~\cite{Jiang2019a}, content feature extraction~\cite{Lei2018}, content clustering~\cite{Hou2018}, and cache replacement~\cite{Wang2017a}. The application of ML techniques in wireless networks and caching is a relatively new research topic. The research community has so far not vigorously focused on applications of ML for cache coding, federated caching, and encoding strategies. Lesser attention has been paid to research issues of security~\cite{Li2018a}, privacy~\cite{Liu2020}, federated learning~\cite{Yu2018}, joint interference alignment and caching~\cite{He2017b}, and cache economics~\cite{Khan2019}. Therefore, many research challenges need to be addressed towards further enhancement of the research domain. In this section, we highlight the key challenges to edge caching based ML techniques. Future research directions for the same are also debated. 

\subsection{Encoding strategies:} 
\begin{figure}
\centering
\includegraphics[width=8.7cm]{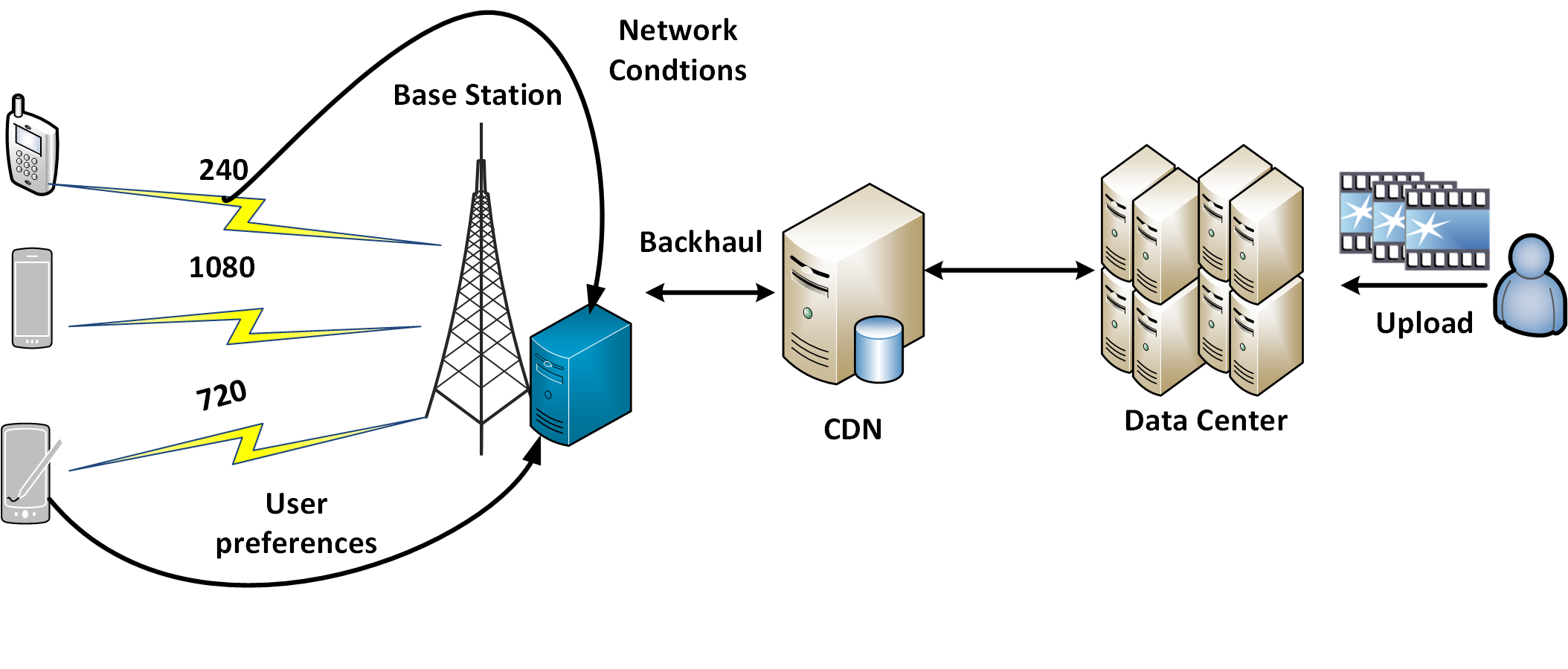} 
\caption {A Framework for Online Transcoding in Edge Networks}
\label{fig:trans}
\end{figure}

Multimedia content is available in a variety of specifications such as, 4K, Quad High Definition, 360$^{o}$, etc. The user preferences and available data rates change the encoding specification from time to time increasing demand on caching capacity~\cite{Paschos2018}. Major content provider like Netflix encode popular videos to more than 120 different bitrates and screen resolutions~\cite{Bilal2019}. Based on the network conditions and user equipment, a different version of same video may be requested, resulting in storage of multiple version of same video in cache, wasting considerable space. Moreover, absence of a specific version means a new request to the content provider is generated resulting in high delay, low QoE, and low CHR. Therefore, instead of requesting a variant of the available cached content, online and in-edge transcoding technique may be applied~\cite{Bilal2018,Bilal2019,Baccour2020a}. The online transcoding of cached content results in considerable gains in storage optimization, users' QoE, access delay, and cost~\cite{Bilal2019}. However, intelligently caching a video representation to avoid redundancy and storage wastage is not explored in detail. Therefore, ML techniques need to be studied to optimize online video caching and transcoding. All of the listed works assumed static encoding parameters ignoring the diversity of user requirements and network conditions. Although work has been done on encoding strategies in wireless edge caching~\cite{He2018,Liao2017,Park2017}, the applications of ML for encoding frameworks are very limited~\cite{Luo2019}. Future research can focus on learning user preferences and network conditions to predict appropriate encoding specifications for pre-fetching content accordingly. Moreover, online transcoding services at the edge of the network can be considered to lower load on transit and core network while providing users with low latency content according to user preferences. Figure~\ref{fig:trans} represents a framework for online transcoding in edge networks with elements for ML where users upload contents to central cloud data centers~\cite{Liu2018a,Bilal2019}. Multiple copies of the same content with different transcoding specifications are stored at the cloud, CDNs, and edge caches. As a result, the capacity of edge cache is limited without the utilization of online transcoding. 

\subsection{Coded Caching:} 
\begin{figure}
\centering
\includegraphics[width=9cm]{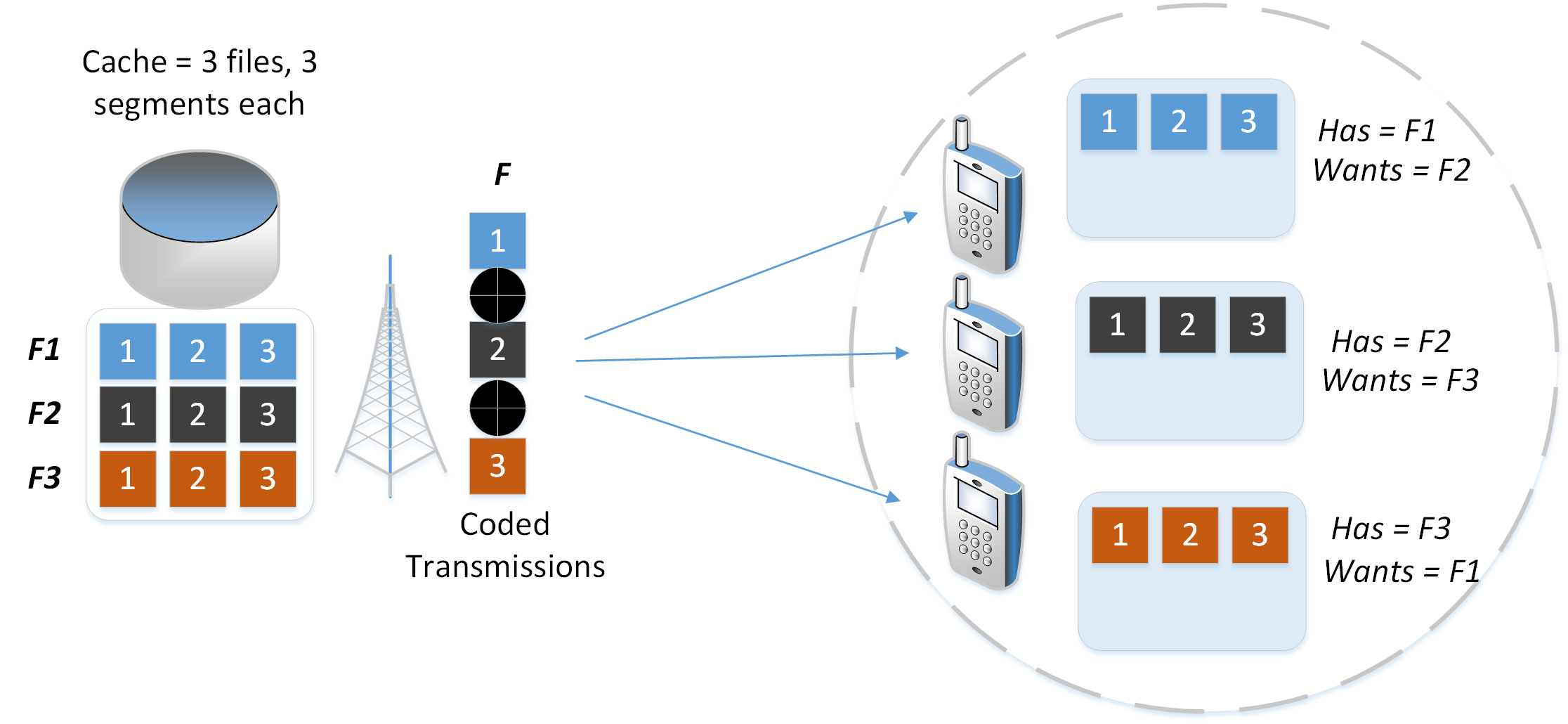} 
\caption {The Concept of Coded Caching in Edge Networks~\cite{Paschos2018}.}
\label{fig:code}
\end{figure}

In coded caching, the content is divided into \textit{wants} (content delivery) and \textit{has} (content placement) parts at each UE. The content placement phase is planned such that future opportunities arise for multi-cast or broadcast delivery of coded content. When the users expose their requests, the server codes the requested content and broadcasts it. Each user decodes its part of content using simple XOR operations and side information~\cite{Xu2019}. A generic framework of coded caching is illustrated in Figure~\ref{fig:code}. Caching coded content at the edge of the network has been demonstrated to substantially increase system throughput and global caching gain~\cite{Chhangte2020,He2019}. Unexplored opportunities exist for the application of ML techniques towards coded caching. The user requests can be predicted with the help of ML techniques to obtain more opportunities of coded content delivery. Moreover, federated feature extraction from the placed content at UEs can help in the distributed prediction of content requests with lower privacy risk. 

\subsection{Federated caching:} Federated/joint caching means that content is cached and shared in the network among multiple MNOs~\cite{Taleb2019}. Federated caching reduces content duplication with a resource sharing mechanism that extends beyond the operational boundary of an MNO~\cite{Fadlullah2020,Qian2020}. However, none of the listed research works has benefited from federated caching which is more optimal than a single MNO caching scenario in terms of cost and resource management. Considerable gains in cost and cache hit ratio may be achieved using federated caching. However, multiple autonomous MNOs have proprietary network equipment and architectures with no sharing and access permissions. Therefore, the central access positions, such as CRANs and virtualized BBU pools offer an optimal place for collaborative and federated caching. However, novel federated caching architectures need to be designed specifically considering privacy and security. Moreover, appropriate cost and monetization models need to be developed considering complex access patterns~\cite{Cui2020,Fadlullah2020}. Currently, there is no consensus on an architecture of federated caching in edge networks that defines cost and payment models for MNOs. Federated caching can help lower the cost of training an ML model by distributing it among multiple MNOs. Moreover, federated caching can lead to better predictions while aggregating data from multiple MNOs. Furthermore, TL techniques can be applied to explore the diffusion of ML outcomes across multiple MNOs. Therefore, a notable future research direction can be federation of MNOs for ML-based caching in edge networks~\cite{Yu2018,Bilal2019}.  

\subsection{Federated learning:} In a general ML approach, data is gathered from users and sent to a central server for feature extraction and training of the ML model. This approach is susceptible to privacy and security issues due to data communication and storage at a central store~\cite{Lim2020,Rehman}. An alternate approach is federated learning which does not require centralized data aggregation and lowers privacy and security risks. Instead of centralized intelligence, individual nodes or users train ML models on local data to collaboratively learn parameters of a distributed ML model~\cite{Zhang2018a,Wang2020b}. The benefits of privacy preservation from federated learning have been investigated in a few of the listed works~\cite{Liu2020,Yu2018}. However, the debate of privacy-accuracy tradeoff expanding from federated learning is yet to be explored in edge caching. Moreover, the above-mentioned works have utilized federated learning for data clustering. Other ML algorithms for popularity prediction, problem optimization, and feature extraction are yet to be investigated with the approach of federated learning.     

\subsection{Delivery strategies:} All of the discussed works have applied explicit D2D cache delivery or implicit unicast delivery. CoMP~\cite{Zhou2017} and multi-cast~\cite{Bilal2019} delivery have not been applied independently to cache delivery. Cache delivery techniques have not been tested in conjunction with the application of ML in edge caching. Multi-cast represents a viable and efficient option when amalgamated with caching and ML for not only RAN but also for transit and backhaul links. Predictive caching based on user preferences may considerably enhance RAN utilization and QoE where the predicted content is multi-cast to multiple UEs simultaneously at non-peak hours and highest bit-rates. Such a multi-cast delivery may occur at any level including \textbf{(a)} MNO core while sending contents to multiple RANs \textbf{(b)} BS level while sending content to multiple UEs within RAN, and \textbf{(c)} D2D level where the content may be multi-cast to multiple UEs in range for future access~\cite{Zahoor2020}. However, very little work has been done in this regard. This point presents a venue for future research works to incorporate CoMP and multi-cast delivery with ML-based caching techniques~\cite{Bilal2019,Wang2017}. Some research areas that may incorporate multi-cast delivery, caching, and ML include \textbf{(a)} community detection by ML techniques and caching at influential users using multi-cast delivery, \textbf{(b)} advertisements and pre-cached content dissemination using multi-cast delivery, and \textbf{(c)} delivery of cost modeling and FL parameters with multi-cast. 

\subsection{User preferences and content popularity:} Some of ML techniques such as TL and RL work with little or no data as input. However, data regarding user preference is required for content clustering and popularity prediction~\cite{Liu2020,Skaperas2020}. The user traffic is https, hence, the application layer content is hidden from the network resources that transport data~\cite{Li2018}. Users can agree to share their preferences, or be monitored for content access at CDNs or destination server~\cite{Lei2018,Din2019}. Therefore, the training data for content request and popularity prediction can only be available from a central server. On the other hand, only 1\% of the Facebook videos account for more than 80\% of the watch time~\cite{Goian2019,Baccour2020a}. Finding this small class of the CDN content popular in an edge community in a certain time interval is a complex but highly beneficial task. The spatial granularity of a BS coverage area and the dynamic number of associated users results in challenging content popularity models~\cite{Liu2016}. Future research works need to explore ML techniques for time-series content popularity with respect to edge communities. 

\subsection{Privacy Issues:} In some use cases, the users may agree to share data about their preferences in order to facilitate ML techniques. In such cases, methods that should be incorporated to assure data privacy and security have not been debated in detail~\cite{Liu2020,He2018,Wang2019b}. For instance, users can share their content preference data to facilitate content clustering (K-means), topic modeling, and social communities identification for optimal caching decisions. Similarly, user mobility data is essential for accurate mobility prediction and UE-BS association models in edge networks. However, the centralized collection of mobility and user preference data is vulnerable to large-scale surveillance. Utilizing user location and preference data will necessitate the establishment of trust protocols among UE and the MNOs~\cite{He2018,Chang2018,Zhang2018a}. Moreover, trust needs to be established among UEs for D2D content sharing with or without the help of a dedicated trust agent~\cite{Zhang2018b,Jiang2019}. Furthermore, security measures should be exercised while the user's data is stored for ML applications at MEC or CDN servers. In the MEC and edge computing scenario, all such protocols should be light-weight so that they do not burden the energy resources of the UE and edge servers extensively~\cite{Khurshid2019,Hussain2019,Adeel2019}. User privacy~\cite{Liu2020} and security~\cite{He2018} issues related to edge caching have been studies. Specifically, work has been done on RL based security in edge caching~\cite{Xiao2018}. However, these works have not been evaluated on desirable characteristics for edge caching, such as light-weight and reliability. Blockchain-based privacy and security solutions have been proposed for edge networks~\cite{Kang2018,Yang2020}. Two research issues need attention regarding the integration of Blockchain services in edge networks: \textbf{(a)} distribution of compute-intensive blockchain workload among edge devices, and \textbf{(b)} monetization/reward policy for collaborative blockchain execution among edge nodes~\cite{Salah2019}.    

\subsection{Cost and monetization models:} Caching economics is one of the least addressed issues in edge caching. Caching at UE and D2D communication leads to the storage, power, and data consumption of mobile devices. All UEs do not cache and share the same amount of content leading to an unfair communication scenario. Therefore, incentive models are necessary for UEs that afford more caching and transmission than other UEs~\cite{Khan2019,Ren2019}. An example of a basic incentive mechanism can be higher download data rates and lower delay for the UE with higher sharing/transmission history~\cite{Jiang2019}. However, most D2D caching techniques lack an incentive mechanism and a standard is necessary for practical UE cache deployments. Moreover, the application of ML techniques for edge caching demands that UE, BSs, and other network entities capable of processing power collaboratively share the workload of compute-intensive learning methods. Based on the distributed contributions, each processing entity should be rewarded for participation in the learning process as all entities receive the same benefits of the optimal caching decisions~\cite{Wang2019a,Debe2020}. Relay nodes may be used in D2D communication, where the sender and receiver may not be in range. Specific monetization models for relay nodes also needs to be defined so that an adhoc setup for data delivery may be established where sender and receiver may reside many hops away. To aggregate issues, compute-intensive Blockchain services are being integrated with edge networks. Despite the many-fold applications of blockchain in edge networks, the distribution of workload and monetization of resources among edge devices remains the main research issue~\cite{Yang2019a,Liu2018b}. 

\subsection{Complexity:} The wireless edge networks have complex dynamics. When considering all dynamic parameters, such as channel variations, user mobility, user preference, content popularity, the optimization of caching cost (data transmission, delivery time, etc), the time complexity of solving optimization problems increases exponentially~\cite{Li2018,Alasmary2019,He2017b}. In particular, the joint optimization of cache placement and physical layer transmission is deemed NP-hard~\cite{Paschos2018}. The caching optimization decisions are also online in nature, where changing dynamics require updates in the solution. Therefore, current research works do not incorporate multiple edge dynamics while assuming some of the parameters as constant to limit the complexity of the mathematical formulation. However, for higher CHR and more applicable solutions, caching decision problems with realistic assumptions are necessary~\cite{Sadeghi2017,Yin2018a}. Suboptimal solutions are seldom considered to reduce the complexity of online solutions for edge caching~\cite{Wang2017b,Shan2019}.  

\subsection{Standard Data Set:} A standard user preference and edge-based content popularity data set is required to compare research works and formulate baseline results~\cite{Sun2019}. Existing research works either predict content popularity from no data or a data set that is not public. Furthermore, the existing data sets derive data from CDN or cloud sources that do not represent an edge network~\cite{Liu2019}. The MovieLens~\cite{Harper2015} data set containing information about real-world movies, users, and ratings are often applied for the evaluation of the recommendation systems. References~\cite{Adeel2019,Thar2018,Yu2018,Song2019,Garg2019} utilize the MovieLens data-set to predict the popularity of the content with various learning techniques. However, the MovieLens data-set represents features for cloud/CDN based user access and can not be mapped to an edge network. Similarly, standard data sets for user mobility also impede baseline research~\cite{Malik2020}. Recent research published a public dataset of Facebook Live videos that can be filtered by location to formulate an edge network representation~\cite{Baccour2020}. The data set contains ID and geo-location among other metadata of more than 1.5 Million Facebook Live videos collected from the Facebook Live Map. Further investigations are required to extract content popularity from the metadata. Moreover, efforts are needed to differentiate and highlight the differences in edge popularity and cloud popularity of the content. Such difference if not considered can significantly effect the cache population and CHR. 

\subsection{Joint network slicing and caching designs:} Network virtualization enabled network slicing is considered a key enabler of NGNs~\cite{Ferrus2018}. Network slicing allows MNOs to manage dedicated logical networks with specific functionality to service user requirements. For example, a vehicular RAN network can be composed of two slices where one supports time-critical driving services while the other network slice serves the bandwidth consuming video streaming services~\cite{Zhang2020}. The distribution of network slices among bandwidth hungry and delay in-tolerant services can guide caching strategies to distribute content among BSs and UE in a RAN network. The joint consideration of network slicing with caching can bring great benefits for the MNO~\cite{Xiang2017}. However, the joint network slicing and caching designs have been evaluated by only one of the listed works~\cite{Xiang2020}. Further work is required to understand the interaction of network slicing with edge caching in edge resource management scenarios.

\subsection{UAV assisted caching:} Researchers~\cite{Chen2017} proposed a novel UAV assisted caching scheme for edge networks. The scheme utilizes mmWave frequency for UAV-to-user communications while deploying UAV in user density areas and clusters. The UAV-to-user channel can provide high SNR than the BS-to-user channel in some scenarios. ML techniques can be applied to predict the content that must be cached in the UAV considering a user cluster that may request specific contents. Moreover, the population and spatial features of the cluster can be predicted by ML techniques for optimal cache decisions and delivery. However, mmWave frequency is highly sensitive to the physical environment and obstacles. Path loss occurs due to user mobility and line-of-sight changes. Coverage and blockage zones change dynamically which requires content duplication and highly dynamic content placement strategy. Therefore, user mobility and content popularity prediction models need to be accurate while accommodating the dynamic behavior of the mmWave channels~\cite{Li2018,Giatsoglou2017}. Future research works need to consider the constraints of mmWave communication while designing optimal D2D caching policies.


\section{Conclusion}\label{sec:conc}
We comprehensively surveyed ML-based edge caching techniques in this article. We examined the role of 5G, SDN, and NFV in edge caching. We provided an exhaustive taxonomy of the research domain with details of background topics including ML, edge networks, and caching. The state-of-the-art ML techniques were categorized into supervised learning, unsupervised learning, RL, NN, and TL based on application to problems of prediction and clustering. We critically compared and analyzed the research work on parameters, such as ML method, network scenario, ML objective, caching objective, caching location, replacement techniques, content popularity distribution, content request distribution, and user mobility models. We further complemented our debate on ML-based edge caching by discussing future challenges and research issues. The research issues and challenges are related to the the exploration of encoding strategies, coded caching, federated caching, initiation of federated learning-based edge caching, multi-cast and CoMP delivery strategies, availability of standard data sets, cost and monetization models for D2D caching, and the establishment of light-weight privacy-preserving and security mechanisms.     
\subsection*{Acknowledgments}
This work is supported by the Postdoctoral Initiative Program at the Ministry of Education, Saudi Arabia.

\bibliography{dc2}

\end{document}